\documentclass[pra,floatfix,showpacs,amsmath,twocolumn]{revtex4}

\usepackage{amssymb}
\usepackage{amsmath}
\usepackage{amsthm}
\usepackage{amstext}
\usepackage{graphicx}
\usepackage{graphics}

\theoremstyle{plain}
\newtheorem{theorem}{Theorem}[section]
\newtheorem{lemma}[theorem]{Lemma}

\theoremstyle{definition}

\newcommand{\bea}{\begin{eqnarray}}
\newcommand{\eea}{\end{eqnarray}}
\def\bi{\begin{itemize}}
\def\ei{\end{itemize}}
\def\bc{\begin{center}}
\def\ec{\end{center}}

\newcommand*{\blambda}{\boldsymbol{\lambda}}

\def\<{\langle}
\def\>{\rangle}

\def\opone{\leavevmode\hbox{\small1\kern-3.8pt\normalsize1}}

\newcommand{\one}{\mbox{$1 \hspace{-1.0mm}  {\bf l}$}}
\def\tr{\mathrm{tr}}
\def\ket#1{\left| #1\right>}
\def\bra#1{\left< #1\right|}

\newcommand{\proj}[1]{\ket{#1}\bra{#1}}

\newcommand*{\bbN}{\mathbb{N}}
\newcommand*{\bbR}{\mathbb{R}}

\newcommand*{\cB}{\mathcal{B}}
\newcommand*{\cC}{\mathcal{C}}
\newcommand*{\cD}{\mathcal{D}}
\newcommand*{\cE}{\mathcal{E}}
\newcommand*{\cF}{\mathcal{F}}
\newcommand*{\cH}{\mathcal{H}}
\newcommand*{\cP}{\mathcal{P}}
\newcommand*{\cS}{\mathcal{S}}

\newcommand*{\cX}{\mathcal{X}}
\newcommand*{\cY}{\mathcal{Y}}
\newcommand*{\cZ}{\mathcal{Z}}

\newcommand*{\eps}{\varepsilon}

\newcommand*{\Hc}{H}
\newcommand*{\Hq}{S}

\newcommand{\lperp}{{\raisebox{-1.2pt}{$\scriptstyle{\perp}$}}}

\newcommand*{\dist}{\delta}

\newcommand*{\QBER}{\mathrm{QBER}}
\newcommand*{\ntot}{n} 
\newcommand*{\aux}{\mathrm{aux}}
\newcommand*{\pe}{\mathrm{pe}}
\newcommand*{\data}{\mathrm{data}}
\newcommand*{\ndata}{n_{\data}} 
\newcommand*{\ndatas}{n}
\newcommand*{\npe}{n_{\pe}} 
\newcommand*{\naux}{n_{\aux}} 
\newcommand*{\bA}{\mathbf{A}}
\newcommand*{\bB}{\mathbf{B}}
\newcommand*{\bS}{\mathbf{S}}
\newcommand*{\bs}{\mathbf{s}}
\newcommand*{\bX}{\mathbf{X}}
\newcommand*{\bx}{\mathbf{x}}
\newcommand*{\bY}{\mathbf{Y}}
\newcommand*{\by}{\mathbf{y}}
\newcommand*{\bU}{\mathbf{U}}
\newcommand*{\bu}{\mathbf{u}}
\newcommand*{\bV}{\mathbf{V}}
\newcommand*{\bv}{\mathbf{v}}
\newcommand*{\bW}{\mathbf{W}}
\newcommand*{\bw}{\mathbf{w}}
\newcommand*{\bZ}{\mathbf{Z}}
\newcommand*{\bZb}{\mathbf{Z'}}
\newcommand*{\cEBell}{\cE_{\mathrm{Bell}}}
\newcommand*{\bn}{\mathbf{n}}

\begin{document}

\title{An information-theoretic security proof for QKD protocols}

\author{R. Renner$^{1}$, N. Gisin$^{2}$, and B. Kraus$^{2}$}
\affiliation{$^1$ Computer Science Department, ETH Z\"urich CH-8092
  Z\"urich, Switzerland\\ $^2$ Group of Applied Physics, University of
  Geneva, CH-1211 Gen\`eve 4, Switzerland }

\begin{abstract}
  
  We present a new technique for proving the security of quantum key
  distribution (QKD) protocols. It is based on direct
  information-theoretic arguments and thus also applies if no
  equivalent entanglement purification scheme can be found.  Using
  this technique, we investigate a general class of QKD protocols with
  one-way classical post-processing. We show that, in order to analyze
  the full security of these protocols, it suffices to consider
  collective attacks.  Indeed, we give new lower and upper bounds on
  the secret-key rate which only involve entropies of two-qubit
  density operators and which are thus easy to compute.  As an
  illustration of our results, we analyze the BB84, the six-state, and
  the B92 protocol with one-way error correction and privacy
  amplification.  Surprisingly, the performance of these protocols is
  increased if one of the parties adds noise to the measurement data
  before the error correction. In particular, this additional noise
  makes the protocols more robust against noise in the quantum
  channel.

\end{abstract}


\maketitle

\section{Introduction}

Classical key distribution schemes can only be secure under strong
assumptions, e.g., that the computing power or the storage capacity of
a potential adversary is limited. In contrast, quantum key
distribution (QKD) allows for provable security under the sole
assumption that the laws of physics are correct. This ultimate
security is certainly one of the main reasons why so much theoretical
and experimental effort is undertaken to investigate QKD protocols
and, in particular, to make them practical~\cite{GiRi02,Idq,MQ}.

One of the most challenging theoretical problems in the context of QKD
is to determine sufficient and/or necessary conditions for the
security of QKD protocols. This is exactly what we are concerned with
in this paper. To be more precise, we investigate the security of a
general class of QKD schemes which includes the most popular ones such
as the BB84, the six-state, and the B92 protocol
\cite{BB84,BeGi99,Be92}.  Our results hold with respect to a model
where two legitimate parties, traditionally called Alice and Bob, are
connected by a quantum channel as well as an authentic, but otherwise
fully insecure, classical channel~\footnote{If Alice and Bob initially
  share a short key, they can use a classical authentication scheme in
  order to implement an authentic channel.}. We assume that Alice's
source as well as Bob's detector are perfect, whereas an adversary
(Eve) might have full control over the quantum channel \footnote{One
  possibility to deal with imperfections of the source or the detector
  is to include them into the model of the quantum channel (where dark
  counts might, e.g., be replaced by random bits). This, however,
  corresponds to a situation where Eve has partial control over these
  devices, which might be unreasonable.}

QKD protocols can usually be divided into a quantum and a classical
part: In the quantum part, the transmitter (Alice) sends qubits (or
more generally, some $d$-level physical systems) prepared in certain
states to the receiver (Bob).  The states of these qubits are
encodings of bit values randomly chosen by Alice. Bob performs a
measurement on the qubits to decode the bit values. For each of the
bits, both the encoding and the decoding are chosen at random from a
certain set of operators.  After the transmission step, Alice and Bob
apply a \emph{sifting} where they publicly compare the encoding and
decoding operator they have used and keep only the bit pairs for which
these operators match.

Once Alice and Bob have correlated bitstrings, they proceed with the
classical part of the protocol. In a first step, called
\emph{parameter-estimation}, they compare the bit-values for a
randomly chosen sample of their strings, which gives an estimate for
the quantum bit error rate (QBER), i.e., the fraction of positions
where Alice and Bob's strings differ.  Note that the QBER is a direct
measure for the secrecy of Alice and Bob's strings, since any
eavesdropping strategy would, according to the laws of quantum
mechanics (no-cloning theorem) perturb the correlations between
them~\footnote{For a fixed attack, the QBER might still take different
  values with certain probabilities. (Note that the average QBER is
  irrelevant in this context.)}. If the QBER is too high, Alice and
Bob decide to abort the protocol.  Otherwise, they apply a
\emph{classical (post)-processing} protocol to distill a secret key,
using either one-way or two-way classical communication.  One-way
post-processing protocols usually consist of \emph{error correction}
and \emph{privacy amplification}~\footnote{Error correction and
  privacy amplification might also be combined into one single
  protocol step.}.  For the error correction, Alice sends certain
information to Bob such that he can reconstruct Alice's string.  Once
Alice and Bob have identical strings, privacy amplification is used to
compute a final key on which the adversary has virtually no
information. We shall see that the performance of such one-way
protocols can generally be increased if Alice additionally applies
some \emph{pre-processing} to her initial string before starting with
the error correction.

Any realistic quantum channel is subject to noise. Consequently, even
in the absence of an adversary Eve, the QBER is non-zero. On the other
hand, Eve might in principle replace the real (noisy) quantum channel
with an ideal noise-free channel and could thus tap mildly into the
quantum communication such as to introduce precisely the original
amount of noise. Hence, when proving the security of a protocol, one
has to assume that all the noise is due to Eve. This raises the
following question: What is the maximum QBER, i.e., the maximum
tolerated channel noise, such that Alice and Bob can still generate a
secure key? Clearly, the answer to this question depends on the amount
of information that Eve might have gained by her attack.

Ideally, one does not want to impose any restriction on Eve's power.
That is, any strategy allowed by the laws of physics has to be
considered. On the other hand, the set of \emph{all} possible attacks
is usually difficult to handle.  In order to cope with these
conflicting objectives, three classes of attacks have been considered.
The smallest class only contains the so-called \emph{individual
  attacks}, where Eve is restricted to interacting with each of the
signal systems sent by Alice separately.  That is, for each of the
signal systems, Eve attaches an auxiliary system and applies some
fixed unitary operation. Finally, Eve measures each of these systems
individually right after the sifting step, i.e., before Alice and Bob
start with the classical processing.  The class of \emph{collective
  attacks} is defined similarly, but the last requirement is dropped.
That is, Eve might wait with her measurement until the very end of the
protocol. In particular, the measurement she chooses might depend on
the messages Alice and Bob exchange for error correction and privacy
amplification.  Moreover, she might measure all her auxiliary systems
jointly.  Not much is known about this class, and research has more
concentrated on the class of \emph{coherent attacks}, which is the
most general one.  In particular, Eve might let all the signal systems
interact with one large auxiliary system, which she only measures at
the very end of the protocol.

Many~\footnote{This is not true for the first security proof of QKD
  against the most general attacks due to Mayers~\cite{Mayers96},
  which is based on different techniques.} of the previous security
proofs of QKD protocols are based on the following observations
\cite{ShPr00,Lo01,GoLo03,TaKoIm03}:
\begin{enumerate}
\item Instead of preparing a system in a certain state and then
  sending it to Bob, Alice can equivalently prepare an entangled
  state, send one of the qubits to Bob, and later measure her
  subsystem.  In doing so, she effectively prepares Bob's system at a
  distance.
\item If the joint system of Alice and Bob is in a pure state, then it
  cannot be entangled with any third party; in particular it cannot be
  entangled with any of Eve's auxiliary systems. Hence, simple
  measurements provide Alice and Bob with data totally oblivious to
  Eve.
\item If furthermore the state shared by Alice and Bob is maximally
  entangled, then their measurement results are maximally correlated.
  Hence, if Alice and Bob performed some entanglement purification
  protocol \cite{Bennett, Gisin}, they would end up with the desired
  secret bits.
\item Since one is interested in the security of protocols implemented
  with nowadays technology, Alice and Bob's operations should not
  require the storage of quantum states, i.e., one does not want them
  to run a general entanglement distillation protocol. To overcome
  this problem, one uses the fact that certain entanglement
  distillation protocols are mathematically equivalent to quantum
  error correction codes. There exists a class of such codes, called
  CSS codes, which have the property that bit errors and phase errors
  can be corrected separately. Since the final key is classical, its
  value does not depend on the phase errors. Hence, Alice and Bob
  actually only have to correct the bit-errors, which is a purely
  classical task.
\end{enumerate} 

This method for proving the security of QKD protocols is very elegant,
but raises two different questions.  First, is the detour via
entanglement purification really necessary? Is it optimal?  Or might
other methods lead to better results?  Secondly, must all
cryptographers learn the intricate theory of entanglement?  Is there
an explanation of the results within the language of information
theory?  As we shall see, the theory of entanglement purification, as
explained above, is not necessary and also too pessimistic (from Alice
and Bob's point of view).

In fact, we present a technique for proving the security of QKD
protocols which does \emph{not} rely on entanglement purification.
Instead, it is based on information-theoretic results on the security
of privacy amplification~\cite{KoMaRe03,RenKoe05}, which have first
been applied in~\cite{ChReEk04} to analyze the security of a generic
QKD protocol similar to the one we are considering here~\footnote{The
  proof technique introduced in~\cite{ChReEk04} applies to most of the
  known protocols with one-way error-correction and privacy
  amplification (but without pre-processing). It is based on the
  result of~\cite{KoMaRe03} and the fact that the rank of a
  purification of Alice's and Bob's system can be bounded.} (see
also~\cite{BenOr02} for a similar approach).  Since secret key
agreement might be possible even if the initial quantum state, the
state Alice and Bob share before error correction and privacy
amplification, does not allow for entanglement purification, our
method generally leads to more optimistic results than any method
based on entanglement purification.  In addition, we show that the
final key is secure according to a so-called \emph{universally
  composable} security definition.  This implies that the key can
safely be used in \emph{any} arbitrary context.  Remarkably, this is
not the case for most of the known security definitions (cf.\ 
discussion in~\cite{RenKoe05}).

One interesting example illustrating the strength of our technique is
the BB84 protocol or the six-state protocol, where, in the classical
processing step, Alice additionally adds some (large) amount of noise
to her measurement data. We show that, surprisingly, this noise
generally increases the rate at which Alice and Bob can generate
secret key bits. However, the density operator of Alice and Bob's
system after the noise has been introduced is not entangled, i.e., any
security proof based on entanglement purification fails.

The paper is organized as follows: In Section~\ref{sec:genprot}, we
describe and analyze a generic QKD protocol using one-way classical
post-processing. According to the discussion above, the protocol is
subdivided into a quantum and a classical part.  In
Section~\ref{sec:dist}, which is devoted to the quantum part, we
review our result presented in \cite{KrGiRe04}. It states that the
density operator describing Alice and Bob's information after the
quantum communication can be considered to be a symmetric (with
respect to permutations of the qubit pairs) Bell-diagonal state.  The
classical part of the protocol is then studied in
Section~\ref{sec:cpp}. Using some recent results of classical and
quantum information theory~\cite{RenWol04,RenKoe05}, we analyze the
performance of the classical post-processing. In
Section~\ref{sec:lowbound}, we combine the main statements of
Sections~\ref{sec:dist} and~\ref{sec:cpp} and derive an expression for
the secret-key rate which only involves a minimization over a certain
set of two-qubit states which correspond to collective attacks. In
Section~\ref{sec:upbound}, we give an upper bound on the secret-key
rate for any protocol with one-way classical post-processing, again
involving only two-qubit density operators.  Finally, in
Section~\ref{sec:examples}, we apply our methods to the BB84, the
six-state, and the B92 protocol. In addition, we show that the
efficiency of each of these protocols can be increased if one of the
parties adds noise to her measurement data.

\section{A general QKD protocol using one-way communication} \label{sec:genprot}

In this section, we describe a general class of QKD protocols
employing one-way classical post-processing.  This class contains the
BB84, the six-state, and the B92 protocol \cite{BB84,BeGi99,Be92},
among many others.  Each of these protocols consists of a quantum and
a classical part: The quantum part includes the distribution and
measurement of quantum information, and is determined by the operators
Alice and Bob use for their encoding and decoding.
Section~\ref{sec:dist} is devoted to the analysis of this part.
Generally speaking, we review our result proven in~\cite{KrGiRe04}
which states that the density operator describing Alice and Bob's
system after the distribution of quantum information can be assumed to
be symmetric (cf.\ equation~\eqref{rhogen}).  Section~\ref{sec:cpp}
deals with the classical part of the QKD protocol, i.e., parameter
estimation and post-processing. We first give a description of a
post-processing scheme and then derive an expression for the maximum
length of the key that this scheme can generate, depending on the
information that Alice and Bob share after the quantum part of the QKD
protocol.

To simplify the presentation of our results, we assume that the
physical systems which Alice sends to Bob are qubits. However, a
generalization to higher dimensions is straightforward.  Throughout
the paper, we use the following notation: Vectors $(l_1,\ldots l_n)$
are denoted by bold letters $\mathbf{l}$. We use capital letters as
subscripts for density operators, e.g., $\sigma_{A B}$, to denote the
subsystems they act on. A bold letter indicates that the corresponding
subsystem is itself a product of many (identical) systems.
Furthermore, for any state $\ket{\Phi}$, $P_{\ket{\Phi}}=\proj{\Phi}$
is the projector onto $\ket{\Phi}$.

\subsection{Quantum part: Distribution of quantum information and measurement} \label{sec:dist}

The quantum part of a QKD protocol is specified by the encoding and
decoding operations employed by Alice and Bob. For the following, we
assume that Alice uses $m$ different encodings, with index $j \in J :=
\{1,\ldots, m\}$. For each $j \in J$, $\ket{\phi^0_j}$ and
$\ket{\phi^1_j}$ denote the states used to encode the bit values $0$
and $1$, respectively.

In the first step of the protocol, Alice randomly chooses $n$ bits
$x_1, \ldots, x_n$ and sends $n$ qubits prepared in the states
$\ket{\phi^{x_1}_{j_1}}, \ldots, \ket{\phi^{x_n}_{j_n}}$ to Bob, for
randomly chosen encodings $j_1, \ldots, j_n$. Upon receiving these
states (which might have undergone some perturbation, possibly caused
by an attack) Bob applies his measurements to obtain classical bits
$(y_1, \ldots, y_n)$.  Finally, Alice and Bob employ a sifting
sub-protocol, where they only keep the qubit pairs for which the
encoding and the measurement operation that they have applied are
compatible.

As demonstrated in~\cite{KrGiRe04}, this protocol can equivalently be
described as a so-called entan\-glement-based scheme~\cite{Ekert91}.
For this purpose, we define the encoding operators $A_j :=
\ket{0}\bra{(\phi^0_j)^\ast}+\ket{1}\bra{(\phi^1_j)^\ast}$ and the
decoding operators
$B_j=\ket{0}\bra{{\phi}_j^1}^\lperp+\ket{1}\bra{\phi_j^0}^\lperp$,
where $\{\ket{0}, \ket{1}\}$ is some orthonormal basis, in the
following called $z$-basis. For $x=0,1$ and $j \in J$,
$\ket{0}\bra{(\phi^0_j)^\ast}$ denotes the complex conjugate of
$\ket{\phi_j^x}$ in the $z$-basis, and $\ket{{\phi}_j^x}^\lperp$ is
some (not necessarily normalized) state orthogonal to
$\ket{\phi_j^x}$.

For the entanglement-based scheme, Alice simply prepares $n$ two-qubit
systems in the state $A_{j_i} \otimes \one \ket{\Phi^+}$, where
$\ket{\Phi^+}=1/\sqrt{2} (\ket{0,0}+\ket{1,1})$, and sends the second
qubit to Bob. Then, Bob randomly applies one of the operators $B_{j}$
to the qubit he receives. Finally, Alice and Bob measure their systems
and associate to the outcome the bit values $0$ or $1$.

The description of a QKD protocol as an entan\-gle\-ment-based scheme
is very convenient for the security analysis. In particular, instead
of considering the quantum communication between Alice and Bob, it
suffices to have a characterization of the quantum state
$\tilde{\rho}_{\bA \bB}^n$ held by Alice and Bob before they apply
their measurements.

Consider now a slight extension of the protocol where Alice and Bob
randomly permute the positions of the measured bit pairs and,
additionally, at each position, flip the values of both bits with
probability one half. In the entanglement based version of the
protocol, these (purely classical) operations can equivalently be
applied to the initial quantum state $\tilde{\rho}_{\bA \bB}^n$ of
Alice and Bob. For the following, we restrict our attention to the
partial state $\tilde{\rho}_{\bA \bB}^{\ndata}$ containing only the
$\ndata$ particle pairs which are later used for the computation of
the final key (but not for parameter estimation) and which are
measured with respect to the $z$-basis~\footnote{We will see in
  Section~\ref{sec:lowbound} that one can always assume that all these
  particle pairs are measured with respect to the same basis.}. (To
keep the notation simple, we write in the following $\ndatas$ instead
of $\ndata$.) The common bit-flip is then described by the quantum
operation $\sigma_x \otimes \sigma_x$.  Moreover, we can assume that
Alice and Bob apply random phase flips $\sigma_z \otimes \sigma_z$ to
their qubit pairs, since these do not change the distribution of the
classical measurement outcomes. The resulting state $\rho_{\bA
  \bB}^{\ndatas}$ of Alice and Bob is thus given by $\rho_{\bA
  \bB}^{\ndatas} = \cD_2^{\otimes
  \ndatas}(\cP_{\ndatas}(\tilde{\rho}_{\bA \bB}^{\ndatas}))$ where the
operator ${\cP}_{\ndatas}$ denotes the completely positive map (CPM)
which symmetrizes the state with respect to permutations of the
$\ndatas$ qubit pairs, and where the CPM $\cD_2$ describes the
operation where both $\sigma_x \otimes \sigma_x$ and $\sigma_z \otimes
\sigma_z$ are applied with probability~$\frac{1}{2}$.  This is
equivalent to the random application of any of the operators $\one
\otimes \one$, $\sigma_x \otimes \sigma_x$, $\sigma_y \otimes
\sigma_y$, or $\sigma_z \otimes \sigma_z$, i.e., $\cD_2$ can be
interpreted as the action of a depolarizing channel transforming any
two-qubit state to a Bell-diagonal state.  Consequently, as shown
in~\cite{KrGiRe04}, $\rho_{\bA \bB}^{\ndatas}$ has the simple form
\begin{equation} \label{rhogen} 
\rho_{\bA \bB}^{\ndatas}
=
\sum_{n_1,n_2,n_3,n_4}^\ndatas
\mu_{n_1,n_2,n_3,n_4} \rho_{n_1,n_2,n_3,n_4} \ .
\end{equation}
In this formula, the sum is taken over all $n_1, n_2, n_3, n_4 \in
\mathbb{N}_0$ satisfying $n_1+n_2+n_3+n_4 = \ndatas$ and
$\mu_{n_1,n_2,n_3,n_4}$ are some (real-valued) non-negative
coefficients. Moreover, $\rho_{n_1,n_2,n_3,n_4}$ is the state of
$\ndatas$ qubit pairs defined by
\begin{equation} \label{rhondef} 
\rho_{n_1,n_2,n_3,n_4} 
:= 
{\cP}_{\ndatas}
  \bigl(P_{\ket{\Phi_1}}^{\otimes n_1}\otimes 
   P_{\ket{\Phi_2}}^{\otimes n_2}\otimes 
   P_{\ket{\Phi_3}}^{\otimes n_3}\otimes
   P_{\ket{\Phi_4}}^{\otimes n_4}\bigr) \ ,
\end{equation}
where $P_{\ket{\Phi_1}} := P_{\ket{\Phi^+}}$, $P_{\ket{\Phi_2}} :=
P_{\ket{\Phi^-}}$, $P_{\ket{\Phi_3}} := P_{\ket{\Psi^+}}$, and
$P_{\ket{\Phi_4}} := P_{\ket{\Psi^-}}$ are projectors onto the Bell
states $\ket{\Phi^{\pm}}=1/\sqrt{2}\ket{0,0}\pm \ket{1,1}$ and
$\ket{\Psi^{\pm}}=1/\sqrt{2}\ket{0,1} \pm \ket{1,0})$.  Note that the
state~\eqref{rhogen} is, independently of the protocol, separable with
respect to the different qubit pairs.

To prove the security of our protocol, we will assume that Eve has the
purification of the state~\eqref{rhogen}, which clearly includes all
the information she possibly can get.  It is explained
in~\cite{KrGiRe04} that, if the encoding operators $A_j$ are unitary,
then this assumption is also tight, i.e., there actually exists an
attack which provides Eve with this purification.

Formula~\eqref{rhogen} is already sufficient to prove our main results
(see Section~\ref{sec:lowbound}). However, to simplify the analysis of
certain protocols, it is often convenient to consider the additional
symmetrization (see~\cite{KrGiRe04}) given by the CPM $\cD_1$ defined
by \bea
\label{D1}{\cal
  D}_1(\rho)=1/N \sum_{j} p_{j} A_{j}\otimes B_{j} (\rho)
A_{j}^\dagger\otimes B_{j}^\dagger \ .  \eea Here $p_{j}\geq 0 $
denotes the probability by which Alice and Bob decide (during the
sifting phase) to keep their bits, if they have applied the operation
$A_{j}\otimes B_{j}$, and $N$ is used for the normalization. All
classical data of Alice and Bob (including the bits used for parameter
estimation) are then given by a measurement of the state
$\cD_2(^{\otimes n}(\cD_1^{\otimes n})(\cP_n(\tilde{\rho}^n_{A B})))$
with respect to the $z$-basis.

\subsection{Classical part: Parameter estimation and classical post-processing} \label{sec:cpp}

This section is devoted to the description and analysis of the
classical part of the QKD protocol. We will use here techniques which
partly have been developed in~\cite{ChReEk04}. Assume that Alice and
Bob already hold strings $\bX=(X_1, \ldots, X_n)$ and $\bY = (Y_1,
\ldots, Y_n)$, respectively, which they have obtained by measuring $n$
particle pairs $\rho_{\bA \bB}^n$ distributed in the first part of the
QKD protocol, as described in Section~\ref{sec:dist}. Their goal is to
generate a secure key pair $(\bS_A, \bS_B)$, using $\bX$ and $\bY$.

The protocol we consider consists of two sub-pro\-to\-cols, called
\emph{parameter estimation} and \emph{classical (post)-processing}.
The main purpose of the parameter estimation sub-protocol is to
estimate the amount of errors that have occurred during the
distribution of the quantum information (see Section~\ref{sec:dist}).
To do this, Alice and Bob compare the measurement outcomes for some
randomly chosen qubit pairs.  If the quantum bit error rate is above a
certain threshold $\QBER$, they decide to abort the protocol.

In order to analyze a given QKD protocol, we need to characterize the
initial states $\rho_{\bA \bB}^n$ for which the protocol does not
abort. Clearly, this characterization depends on the threshold
$\QBER$. Let $\Gamma$ be the set of all two-qubit states $\sigma_{A
  B}$ which correspond to a collective attack, meaning that there
exists an operation of Eve such that $\rho_{\bA \bB}^n = \sigma_{A
  B}^{\otimes n}$. The set $\Gamma_\QBER$ is then defined as the
subset of $\Gamma$ containing all states $\sigma_{A B}$ for which the
protocol does \emph{not} abort (with probability almost one). In other
words, if $\sigma_{A B} \in \Gamma_\QBER$, then the protocol is
supposed to compute a secret key when starting with $\rho_{\bA \bB}^n
= \sigma_{A B}^{\otimes n}$. We will see in Section~\ref{sec:lowbound}
that the characterization of the set $\Gamma_\QBER$ is sufficient to
compute lower bounds on the secret-key rate.

After the parameter estimation, if the estimate for the QBER is below
the threshold, Alice and Bob proceed with a classical sub-protocol in
order to turn their only partially secure strings $\bX$ and $\bY$ into
a highly secure key pair $(\bS_A, \bS_B)$. The protocol we consider is
one-way, i.e., only communication from Alice to Bob is needed. It
consists of three steps: \bi 
\item[I)] \emph{Pre-processing:} Using her bit string $\bX$, Alice
  computes two strings $\bU$ and $\bV$, according to some channels
  $\bU \leftarrow \bX$ and $\bV \leftarrow \bU$, defined by
  conditional probability distributions $P_{\bU|\bX}$ and
  $P_{\bV|\bU}$, respectively.  She keeps $\bU$ and sends $\bV$ to
  Bob.  (We will see that, for most protocols, the performance highly
  depends on a clever choice of $\bU$, whereas the string $\bV$ is
  usually not needed.)
\item[II)] \emph{Information reconciliation:} Alice sends error
  correction information $\bW$ on $\bU$ to Bob.  Using $\bY$, $\bV$,
  and $\bW$, Bob computes a guess $\hat{\bU}$ for $\bU$.
\item[III)] \emph{Privacy amplification:} Alice randomly chooses a
  function $F$ from a family of two-universal hash
  functions~\footnote{For a definition and constructions of
    two-universal hash functions, see, e.g., \cite{CarWeg79}
    or~\cite{WegCar81}.} and sends a description of $F$ to Bob.  Then
  Alice and Bob compute their keys, $\bS_A = F(\bU)$ and $\bS_B =
  F(\hat{\bU})$, respectively. \ei

  Before starting with the analysis of this protocol, let us introduce
  some notation. It is most convenient to describe the classical
  information of Alice and Bob as well as the quantum information of
  the adversary Eve by a tripartite density operator $\rho_{\bX \bY
    E}$ of the form
\begin{equation} \label{eq:rhonclass}
  \rho^n_{\bX \bY E}
=
  \sum_{\bx, \by} P_{\bX \bY}(\bx,\by)
    P_{\ket{\bx}} \otimes P_{\ket{\by}} \otimes \rho^{\bx,\by}_E
\end{equation}
where $\{\ket{\bx}\}_\bx$ and $\{\ket{\by}\}_\by$ are families of
orthonormal vectors and where $\rho^{\bx,\by}_E$ is the quantum state
of Eve given that Alice and Bob's random variables $\bX$ and $\bY$
take the values $\bx$ and $\by$, respectively. Similarly, the
classical key pair $(\bS_A, \bS_B)$ together with the adversary's
information $\rho_{E'}^{\bs_A, \bs_B}$ after the protocol execution is
described by a quantum state $\rho_{\bS_A \bS_B E'}$.  We say that
$(\bS_A, \bS_B)$ is \emph{$\eps$-secure} (with respect to $\rho_{E'}$)
if
\begin{equation} \label{eq:epssecure}
  \delta\bigl( \rho_{\bS_A \bS_B E'},
    \sum_{\bs \in \cS} P_{\bS}(\bs)
      P_{\ket{\bs}} \otimes P_{\ket{\bs}} \otimes \rho_{E'}
  \bigr)
\leq
  \eps
\end{equation}
where $P_{\bS}$ is the uniform distribution over all possible keys
$\bs$ and where $\delta(\cdot, \cdot)$ denotes the trace-distance.
This implies that the state $\rho_{\bS_A \bS_B E'}$ describing the key
of Alice and Bob together with the adversary's quantum system is close
to a state where the adversary's system is completely independent of
the key.

The goal of the remaining part of this section is to derive an
expression for the number $\ell^n$ of $\eps$-secure key bits that can
be generated by the above protocol, for an optimal choice of the
protocol parameters. For this purpose, we first consider some fixed
pre-processing, specified by the channels $\bU \leftarrow \bX$ and
$\bV \leftarrow \bU$, for which we compute the maximum key length
$\ell^{\ntot}_{\bU\leftarrow \bX,\bV \leftarrow \bU}$. The quantity
$\ell^n$ is then obtained by optimizing over all choices of the
pre-processing.

Our result is formulated in terms of an information-theoretic
quantity, called \emph{smooth R\'enyi entropy}~\cite{RenWol04} (see
Appendix~\ref{sec:smooth} for more details). Similarly to the Shannon
entropy $H(X)$, the smooth R\'enyi entropy of a random variable $X$,
denoted by $H_\alpha^\eps(X)$, is a measure for the uncertainty about
the value of $X$.  We will also need an extension of this entropy
measure to quantum states.  Similarly to the von Neumann entropy
$S(\rho)$, the smooth R\'enyi entropy $S_\alpha^\eps(\rho)$ of a state
$\rho$ quantifies the amount of randomness contained in $\rho$.

The main ingredient needed for the following derivation is a recent
result on the security of privacy amplification~\cite{RenKoe05} (see
Lemma~\ref{lem:pa}).  Generally speaking, it says that the length of
the key that can be extracted from a string $\bU$ held by both Alice
and Bob is given by the uncertainty of the adversary about $\bU$,
measured in terms of smooth R\'enyi entropies. Applied to the last
step of our protocol, we get
\begin{equation} \label{eq:pa}
  \ell^{\ntot}_{\bU \leftarrow \bX,\bV \leftarrow \bU}
\approx
  S_2^{\eps}(\rho^n_{\bU \bV \bW E}) - S_0^{\eps}(\rho^n_{\bV \bW E}) \ ,
\end{equation}
where $\eps$ depends on the desired security of the final key and
where the approximation ``$\approx$'' means that equality holds up to
some small additive term of the order $O(\log(1/\eps))$. In this
formula, $\rho^n_{\bU \bV \bW E}$ is the density operator describing
the strings $\bU$, $\bV$, and $\bW$, together with the adversary's
knowledge, i.e.,
\begin{multline*}
  \rho^n_{\bU \bV \bW E} \\
= \! \!
  \sum_{\bx, \by, \bu, \bv, \bw} \! \!
    P_{\bX \bY \bU \bV \bW}(\bx, \by, \bu, \bv, \bw) \,
    P_{\ket{u}} \otimes P_{\ket{v}} \otimes P_{\ket{w}}
    \otimes \rho_E^{\bx, \by}
\end{multline*}
where $\{\ket{\bu}\}_\bu$, $\{\ket{\bv}\}_\bv$, and
$\{\ket{\bw}\}_\bw$ are families of orthonormal vectors. Note that,
since the channel connecting Alice and Bob might be arbitrarily
insecure, the key must be secure even if the adversary knows $\bV$ and
$\bW$.

In the next step, we will eliminate the dependence on $\bW$
in~\eqref{eq:pa}. For this, we consider the amount $m$ of (useful)
information contained in $\bW$. Since $\bW$ is needed by Bob in order
to guess $\bU$, $m$ depends on his uncertainty about $\bU$.  In fact,
if an optimal error correction code is applied, then $m$ is roughly
equal to the entropy of $\bU$ conditioned on Bob's information $\bY$
and $\bV$.  More precisely, using Lemma~\ref{lem:ir} described in
Appendix~\ref{sec:known}, we have $m \approx H_0^\eps(\bU|\bY \bV)$.
Hence, when omitting $\bW$ on the right hand side of~\eqref{eq:pa},
the smooth R\'enyi entropies cannot decrease by more than $m$ (see
Appendix~\ref{sec:smooth} for a summary of the properties of smooth
R\'enyi entropy). We thus immediately obtain
\begin{equation} \label{eq:genboundfix}
  \ell^{\ntot}_{\bU \leftarrow \bX ,\bV \leftarrow \bU}
\approx
    S_2^\eps(\rho^{\ntot}_{\bU \bV E}) - S_0^\eps(\rho^{\ntot}_{\bV E})
    - H_0^\eps(\bU|\bV \bY) \ .
\end{equation}

Since the channels $\bU \leftarrow \bX$ and $\bV \leftarrow \bU$
applied by Alice in the first step of the classical post-processing
protocol are arbitrary, we can optimize over all choices of such
channels. We thus conclude that the number $\ell^{\ntot}$ of key bits
that can be generated by the described protocol, for an optimal choice
of all the parameters, is given by
\begin{equation} \label{eq:genbound}
  \ell^{\ntot}
\approx
  \sup_{\substack{\bU \leftarrow \bX \\ \bV \leftarrow \bU }}
    S_2^\eps(\rho^{\ntot}_{\bU \bV E}) - S_0^\eps(\rho^{\ntot}_{\bV E})
    - H_0^\eps(\bU|\bV \bY) \ .
\end{equation}

In the following, we will often consider protocols where the strings
$\bU$ and $\bV$ are computed bit-wise from the string $\bX$.  The
maximum length of the secret key that can be generated by such a
protocol is then given by an expression similar
to~\eqref{eq:genbound}, but where the supremum is only taken over
bit-wise channels $U \leftarrow X$ and $V \leftarrow U$.

\section{A lower bound on the secret-key rate}
\label{sec:lowbound}

The goal of this section is to derive a lower bound for the secret-key
rate which only involves two-qubit states and which is thus easy to
compute. For this purpose, we use the general
expression~\eqref{eq:genbound} of Section~\ref{sec:cpp} for the number
of key bits that can be generated from a given state, together with
the fact that, after symmetrization, any state of Alice and Bob has
the simple form~\eqref{rhogen}.

Let us start with a description of our main result. Consider the QKD
protocol described in Section~\ref{sec:genprot}, where we assume that
Alice uses bit-wise channels $U \leftarrow X$ and $V \leftarrow U$ to
compute $\bU = (U_1, \ldots, U_n)$ and $\bV = (V_1, \ldots, V_n)$,
respectively, from her data $\bX = (X_1, \ldots, X_n)$.  Let
$\Gamma_\QBER$ be the set of two-qubit density operators $\sigma_{A
  B}$ defined in Section~\ref{sec:cpp}, i.e., the protocol aborts
(with high probability) whenever it starts with a product state
$(\sigma_{A B})^{\otimes \ntot}$ for any $\sigma_{A B} \notin
\Gamma_\QBER$. We show that, for an optimal choice of the parameters,
the protocol of the previous section, generates secret key bits at
rate $r := \lim_{\ntot \to \infty} \frac{\ell^{\ntot}}{\ntot}$ where
\begin{equation} \label{eq:singlebound}
  r
\geq
  \sup_{\substack{U \leftarrow X \\ V \leftarrow U}} 
    \, \inf_{\sigma_{A B} \in \Gamma_\QBER}
    \bigl(S(U | V E) - H(U|Y V) \bigr)
  \ .
\end{equation}
In this formula, $S(U | V E)$ denotes the von Neumann entropy of $U$
conditioned on $V$ and Eve's initial information, i.e., $S(U|V E) :=
S(\sigma_{U V E}) - S(\sigma_{V E})$. The state $\sigma_{U V E}$ is
obtained from $\sigma_{A B}$ by taking a purification $\sigma_{A B E}$
of the Bell diagonal state $\sigma_{A
  B}^\mathrm{diag}:=\cD_2(\sigma_{A B})$~\footnote{This means that
  $\sigma_{A B}^\mathrm{diag}$ has the same diagonal entries as
  $\sigma_{A B}$ with respect to the Bell basis.} and applying the
measurement of Alice followed by the classical channels $U \leftarrow
X$ and $V \leftarrow U$.  Similarly, $Y$ is the outcome of Bob's
measurement applied to the second subsystem of $\sigma_{A B E}$.

As~\eqref{eq:singlebound} involves a minimization over the set
$\Gamma_\QBER$ of two-qubit states, our lower bound on the secret-key
rate only depends on the set of possible collective attacks. On the
other hand, the security we prove holds against any arbitrary coherent
attack.  Note also that the statement extends to the situation where
Alice---instead of applying a bit-wise pre-processing on each of the
$n$ bits---uses some operation involving larger blocks, say of length
$m$.  In this case, one has to consider all attacks $U^{\otimes r}$
where the adversary applies the same operation $U$ on each of the $r =
\frac{m}{n}$ blocks.

A crucial task when computing explicit values
for~\eqref{eq:singlebound} is to characterize the set $\Gamma_\QBER$,
This set is determined by the conditions under which the protocol
aborts.  In Section~\ref{sec:examples}, we will demonstrate how
formula~\eqref{eq:singlebound} is computed for concrete QKD schemes
such as the BB84 or the six-state protocol. It turns out that, in
these examples, the maximum is taken if $V \leftarrow U$ is the
trivial channel where $V$ is independent of $U$, i.e., the random
variable $V$ can be omitted.

One method to further reduce the number of parameters is to consider
the set $\cD_2(\cD_1(\Gamma_\QBER))$, which only contains normalized
two-qubit density operators of the form \bea \label{rho1}
\rho^1[\blambda]=\lambda_1 P_{\ket{\Phi^+}}+\lambda_2
P_{\ket{\Phi^-}}+\lambda_3 P_{\ket{\Psi^+}}+\lambda_4
P_{\ket{\Psi^-}}, \eea i.e., Eq.~(\ref{rhogen}) for $n=1$. As
mentioned in Section~\ref{sec:dist} (see~\cite{KrGiRe04} for details),
the state shared by Alice and Bob is---independently of the considered
protocol---measured with respect to the $z$-basis. Hence, we obtain
for the QBER $Q$, computed as an average over the different encodings,
$Q=\lambda_3+\lambda_4$.  Apart from that, the state must be
normalized, which implies that, for any given value of $Q$, there are
at most two free parameters, $\lambda_2$, and $\lambda_3$, i.e.,
$\lambda_1=1-Q-\lambda_2, \lambda_4=Q-\lambda_3$.

To prove~\eqref{eq:singlebound}, we will make use of a known
result~\cite{ChReEk04} on the relation between the statistics obtained
when applying two different measurements $\cE$ and $\cF$ on the
individual subsystems of a symmetric $n$-partite state $\rho^{n}$
(cf.\ Lemma~\ref{lem:Ekert} in Appendix~\ref{sec:known}). Let $\bZ =
(Z_1, \ldots, Z_{k})$ be the outcomes when applying $\cE$ to each of
the first $k$ subsystems of $\rho^n$, for $k \leq n$, and let
$Q_{\bZ}$ be the frequency distribution of the symbols in the string
$\bZ$, i.e., for any possible measurement outcome $z$,
\[
Q_{\bZ}(z) := \frac{|\{i: Z_i = z\}|}{k} \ .
\]
Similarly, let $Q_{\bZb}$ be the frequency distribution of the outcomes
$\bZb = (Z'_1, \ldots, Z'_{k'})$ of $\cF$ applied to $k'$ of the
remaining $n-k$ subsystems of $\rho^{n}$. Lemma~\ref{lem:Ekert}
implies that, if $k$ and $k'$ are large enough, then, with probability
almost one, there exists a density operator $\sigma$ on one subsystem
which is compatible with both of these statistics. Formally, this
means that $Q_{\bZ} \approx P_\cE[\sigma]$ and $Q_{\bZ} \approx
P_\cF[\sigma]$, where $P_\cE[\sigma]$ and $P_\cF[\sigma]$ denote the
probability distributions of the outcomes when measuring $\sigma$ with
respect to $\cE$ and $\cF$, respectively. Moreover, the state $\sigma$
is contained in a certain set $\cB$ which, roughly speaking, contains
all density operators which correspond to the state of one single
subsystem of $\rho^n$, conditioned on any measurement on the remaining
subsystems.

We are now ready to prove expression~\eqref{eq:singlebound} for the
secret-key rate. As in Section~\ref{sec:dist}, we consider an
extension of the protocol where, before invoking the classical part of
the QKD protocol, Alice and Bob symmetrize their strings $\bX$ and
$\bY$. More concretely, they both apply the \emph{same} randomly
chosen permutation on their strings.  Clearly, this is equivalent to a
protocol where Alice and Bob first permute and then measure their bits
(see Section~\ref{sec:dist}).  The state $\rho_{\bA \bB}^{\ntot}$ of
Alice and Bob's system before the measurement is then symmetric. We
can thus assume without loss of generality that the first $\npe$ qubit
pairs are used for the parameter estimation, while the actual key is
generated from the measurement outcomes obtained from the next
$\ndata$ pairs.


Consider now some fixed protocol where the pre-processing is defined
by the channels $U \leftarrow X$ and $V \leftarrow U$. We show that
this protocol is secure as long as the rate at which the key is
generated is not larger than
\begin{equation} \label{eq:singboundfix}
  r_{U \leftarrow X, V \leftarrow U}
=
  \inf_{\sigma_{A B} \in \Gamma_\QBER}  \bigl(S(U | V E) - H(U|Y V) \bigr) \ .
\end{equation}
In other words, $r_{U \leftarrow X, V \leftarrow U}$ is the rate that
can be achieved if the channels $U \leftarrow X$ and $V \leftarrow U$
are used for the pre-processing.  The assertion~\eqref{eq:singlebound}
then follows by optimizing over all channels for the pre-processing.

The proof of~\eqref{eq:singboundfix} is subdivided into two parts. In
the first part, we show that the parameter estimation works correctly,
i.e., if the adversary introduces too much noise, then the protocol
aborts. The second part of the proof is concerned with the security of
the classical post-processing step, that is, if the noise is below a
certain level, then the final key is secure.

For this analysis, we need to consider the state $\rho_{\bA \bB}^{\npe
  + \ndata}$ of the qubit-pairs used for parameter estimation and
classical post-processing. However, in order to simplify the
presentation of the proof, we assume that there is a small number
$\naux:=\ntot-\npe-\ndata>0$ of additional two-qubit pairs which are
not used by the protocol \footnote{If this is not the case, one can
  always change the protocol such that some of the data bits are
  discarded, without reducing its rate.}. In order to get some
information about the structure of the state $\rho_{\bA \bB}^{\npe +
  \ndata}$, we consider a measurement $\cEBell$ with respect to the
Bell basis applied to each of the remaining $\naux$ positions of
$\rho_{\bA \bB}^{\ntot}$. We then analyze the security of our QKD
protocol conditioned on the statistics $Q_{\bW}$ of the outcomes $\bW
= (W_1, \ldots, W_{\naux})$ of this measurement. We show that the
protocol is secure for all values of $Q_{\bW}$, which implies that the
protocol is secure in general (with probability almost one).

Formally, let $P_{\cEBell}[\Gamma_\QBER]$ be the set of probability
distributions obtained by measuring the states $\sigma_{A B} \in
\Gamma_\QBER$ with respect to the Bell basis. We prove the following
two statements:
\begin{enumerate}
\item \label{sec:pe} If $Q_{\bW} \notin P_{\cEBell}[\Gamma_\QBER]$
  then the protocol aborts after the parameter estimation, i.e., no
  key is generated.
\item \label{sec:class} If $Q_{\bW} \in P_{\cEBell}[\Gamma_\QBER]$
  then the key generated by the classical post-processing is secure.
\end{enumerate}

To prove statement~\ref{sec:pe}, let $\cF$ be the measurement that
Alice and Bob apply to each of the $\npe$ qubit pairs used for
parameter estimation and let $Q_{\pe}$ be the frequency distribution
of the measurement outcomes of $\cF$. Since the state $\rho_{\bA
  \bB}^{\ntot}$ is symmetric, we can apply Lemma~\ref{lem:Ekert}
described above, where $\cB$ is defined by the set $\Gamma$ of all
two-qubit states characterizing the collective attacks of Eve, as
described in Section~\ref{sec:cpp}. Hence, there exists a state
$\sigma_{A B} \in \Gamma$ (of a \emph{single} qubit pair) which is
compatible with both the statistics $Q_{\pe}$ and $Q_{\bW}$, i.e.,
$P_{\cF}[\sigma_{A B}] \approx Q_{\pe}$ and $P_{\cEBell}[\sigma_{A B}]
\approx Q_{\bW}$.  Assume now that $Q_{\bW} \notin
P_{\cEBell}[\Gamma_\QBER]$. Because of $P_{\cEBell}[\sigma_{A B}]
\approx Q_{\bW}$, this implies that $\sigma_{A B} \notin
\Gamma_{\QBER}$. On the other hand, since $P_{\cF}[\sigma_{A B}]
\approx Q_{\pe}$, the statistics $Q_{\pe}$ corresponds to the
frequency distribution obtained when measuring each of the $\npe$
subsystems of the product state $(\sigma_{A B})^{\otimes \npe}$ with
respect to $\cF$. Hence, by the definition of the set
$\Gamma_{\QBER}$, the protocol aborts.

We proceed with the proof of statement~\ref{sec:class}. For any
frequency distribution $Q$, let $\rho_{\bA \bB|Q_\bW = Q}^{\ndata}$ be
the state of the $\ndata$ qubit pairs used for generating the final
key, conditioned on the event that the statistics of the measurement
outcomes of the $\naux$ auxiliary pairs is equal to~$Q$. Assume now
that Alice and Bob measure their data bits according to one fixed
basis~\footnote{If the data bits are measured with respect to
  different bases, the argument must be repeated for each basis. This
  is, however, usually not needed. In fact, for an optimal performance
  of the protocol, one of the encodings should be chosen with
  probability almost $1$ whereas the other encodings should only be
  chosen with some small probability~\cite{LoChAr00}. (The bit pairs
  resulting from the latter are then only used for parameter
  estimation.)  This reduces the number of qubit-pairs lost in the
  sifting step.}, called $z$-basis, and, additionally, apply common
random bit-flips.  Then, according to the discussion in
Section~\ref{sec:dist}, it is sufficient to consider states of the
form~\eqref{rhogen}.  In particular, the conditional state $\rho_{\bA
  \bB|Q_\bW = Q}^{\ndata}$ can be written as
\begin{equation} \label{eq:apstate}
  \rho_{\bA \bB|Q_\bW = Q}^{\ndata}
=
  \sum_{n_1,n_2,n_3,n_4}
    \mu_{n_1,n_2,n_3,n_4} \rho_{n_1,n_2,n_3,n_4}
\end{equation}
where $\rho_{n_1,n_2,n_3,n_4}$ is defined by~\eqref{rhondef}. Hence,
if we applied the Bell measurement $\cEBell$ to each of the $\ndata$
subsystems, then, for any $4$-tuple $(n_1, n_2, n_3, n_4)$, with
probability $\mu_{n_1, n_2, n_3, n_4}$, the resulting frequency
distribution $Q_{\data}$ would be equal to $Q_{n_1, n_2, n_3,
  n_4}:=(\frac{n_1}{n}, \frac{n_2}{n}, \frac{n_3}{n}, \frac{n_4}{n})$.
On the other hand, it follows directly from Lemma~\ref{lem:Ekert}
(with $\cE = \cF = \cEBell$) that $Q_{\data} \approx Q_{\bW}$ holds
with probability almost one.  Hence, the coefficients
$\mu_{n_1,n_2,n_3,n_4}$ can only be non-negligible if $Q_{n_1, n_2,
  n_3, n_4}$ is close to $Q_{\bW}$, that is, we can restrict the sum
in~\eqref{eq:apstate} to values $(n_1, n_2, n_3, n_4)$ such that
$Q_{n_1, n_2, n_3, n_4} \approx Q$.

Consider now the product state $(\sigma_{A B})^{\otimes \ndata}$,
where $\sigma_{A B} := \rho^1[Q]$ is the two-qubit state depending on
$Q$ as defined by~\eqref{rho1}. Since the state $(\sigma_{A
  B})^{\otimes \ndata}$ is symmetric, we can also write it in the
form~\eqref{eq:apstate}, with some coefficients
$\mu'_{n_1,n_2,n_3,n_4}$. Again, these coefficients can only be
non-negligible if $Q_{n_1, n_2, n_3, n_4}$ is close to $Q$. Hence, the
states $\rho_{\bA \bB|Q_\bW = Q}^{\ndata}$ and $(\sigma_{A
  B})^{\otimes \ndata}$ have the same structure~\eqref{eq:apstate}
where the coefficients $\mu_{n_1,n_2,n_3,n_4}$ and
$\mu'_{n_1,n_2,n_3,n_4}$ are negligible except for $Q_{n_1, n_2, n_3,
  n_4} \approx Q$. Using this fact, it is a consequence of the results
presented in Appendix~\ref{app:symstates} that the smooth R\'enyi
entropies of the states derived from $\rho_{\bA \bB|Q_\bW =
  Q}^{\ndata}$ are roughly equal to the corresponding entropies of the
states derived from $(\sigma_{A B})^{\otimes \ndata}$. To make this a
bit more precise, let $\rho_{\bU \bV E|Q_\bW=Q}^{\ndata}$ be the state
obtained when applying the measurement of Alice followed by the
channels $U \leftarrow X$ and $V \leftarrow U$ to each of the
subsystems of a purification of $\rho_{\bA \bB|Q_\bW = Q}^{\ndata}$.
Then, Lemma~\ref{lem:Hqcalc} implies that
\[
  S_2^\eps(\rho_{\bU \bV E|Q_\bW=Q}^{\ndata})
\gtrapprox
  \ndata S(\sigma_{U V E})
\]
and
\[
  S_0^\eps(\rho_{\bV E|Q_\bW=Q}^{\ndata})
\lessapprox
  \ndata S(\sigma_{V E})
\]
where $\sigma_{U V E}$ is the state obtained from $\sigma_{A B} :=
\rho^1[Q]$, as described after~\eqref{eq:singlebound}.

Using these identities, it follows from~\eqref{eq:genboundfix} that
the final key generated by the protocol of the previous section, for
fixed channels $U \leftarrow X$ and $V \leftarrow U$, is secure as
long as its length is not larger than
\[
  \ell_{U \leftarrow X, V \leftarrow U}[\sigma_{A B}]
\approx
  \ndata \bigl( S(\sigma_{U V E}) - S(\sigma_{V E}) - H(U|V Y) \bigr) \ ,
\]
for $\sigma_{A B} = \rho^1[Q]$. In other words, $\ell_{U \leftarrow X,
  V \leftarrow U}[\sigma_{A B}]$ is the length of a secure key that
can be extracted when applying the protocol to a state of the form
$\rho_{\bA \bB|Q_\bW = Q}^{\ndata}$.

Since the final key must be secure for all possible initial states for
which the protocol does not abort, we have to take the minimum of this
quantity over the states $\sigma_{A B} = \rho^1[Q]$, for any $Q \in
P_{\cEBell}[\Gamma_\QBER]$.  Since, according to~\eqref{rho1},
$\rho^1[Q]$ is diagonal, the minimum ranges over all diagonal states
$\sigma_{A B}^{\mathrm{diag}}$ whose diagonal elements correspond to
$Q \in P_{\cEBell}(\Gamma_\QBER)$. This is equivalent to say that the
diagonal elements of $\sigma_{A B}^{\mathrm{diag}}$ are equal to the
diagonal entries of a density operator $\sigma_{A B} \in
\Gamma_\QBER$, i.e., the number $\ell$ of key bits generated by the
protocol is given by
\[
  \ell_{U \leftarrow X, V \leftarrow U}
:=
  \inf_{\sigma_{A B} \in \Gamma_\QBER}
    \ell_{U \leftarrow X, V \leftarrow U}[\sigma_{A B}^{\mathrm{diag}}] \ ,
\]
where $\sigma_{A B}^{\mathrm{diag}} := \cD_2(\sigma_{A B})$.  This
concludes the proof of~\eqref{eq:singboundfix} and thus
also~\eqref{eq:singlebound}.

\section{An upper bound on the secret-key rate} \label{sec:upbound}

As demonstrated in Section~\ref{sec:lowbound}, the rate of a QKD
protocol is lower bounded by an expression which only involves von
Neumann entropies of states of single qubit pairs
(cf.~\eqref{eq:singlebound}). In the following, we show that, roughly
speaking, the right hand side of~\eqref{eq:singlebound} is also an
upper bound on the rate if the supremum is taken over all
\emph{quantum} channels (instead of only classical channels) $U
\leftarrow X$ and $V \leftarrow X$.

Clearly, in order to prove upper bounds, it is sufficient to consider
collective attacks. We thus assume that the overall state $\rho_{\bA
  \bB E}^n$ of Alice's, Bob's, and Eve's quantum system has product
form, i.e., $\rho_{A B E}^n = \sigma_{A B E}^{\otimes n}$, for some
tripartite state $\sigma_{A B E}$. Hence, before starting with the
classical processing, the situation is fully specified by the $n$-fold
product state $\sigma_{X Y E}^{\otimes n}$, where $\sigma_{X Y E}$ is
the state obtained when applying Alice's and Bob's measurements to
$\sigma_{A B E}$. Similarly to~\eqref{eq:rhonclass}, $\sigma_{X Y E}$
can be written as
\[
  \sigma_{X Y E}
=
  \sum_{x,y} P_{X Y}(x,y)
    P_{\ket{x}} \otimes P_{\ket{y}} \otimes \sigma^{x,y}_E \ .
\]

We show that the rate $r(\sigma_{X Y E})$ at which secret key bits can
be generated from this situation, using only a public communication
channel from Alice and Bob, is upper bounded by
\begin{equation} \label{eq:singleupbound}
  r(\sigma_{X Y E})
\leq
  \sup_{\substack{\sigma_U \leftarrow X \\ \sigma_V \leftarrow X}} 
    \bigl(S(U | V E)) - S(\sigma_{U | Y V}) \bigr) \ .
\end{equation}
In this formula, the supremum is taken over all density operators
$\sigma_U^x$ and $\sigma_V^x$ depending on $x$. The density operators
occurring in the entropies are then given by the appropriate traces of
\begin{equation} \label{eq:upboundsigma}
  \sigma_{U V Y E}
:=
  \sum_{x, y} P_{X Y}(x,y) \, \sigma_U^x \otimes \sigma_V^x
    \otimes P_{\ket{y}} \otimes \sigma_E^x \ .
\end{equation}

A similar upper bound for the key rate follows from a result of
Devetak and Winter~\cite{DevWin03}. In contrast
to~\eqref{eq:singleupbound}, their formula involves an additional
limes over the number $n$ of product states, whereas the supremum only
involves classical channels $U \leftarrow X$ and $V \leftarrow U$.

Because of the optimization over the density operators $\sigma_U^x$
and $\sigma_V^x$, expression~\eqref{eq:singleupbound} is generally
hard to evaluate. To simplify this computation, it is convenient to
consider measurements of Eve, resulting in classical values $Z$. In
this case, the bound corresponds to a known result due to Csisz\'ar
and K\"orner~\cite{CsiKoe78},
\begin{equation} \label{eq:upboundcl}
  r(X,Y,Z)
=
  \sup_{\substack{U \leftarrow X \\ V \leftarrow U}} 
    \bigl(H(U | V Z)) - H(U | Y V) \bigr) \ .
\end{equation}

The proof of the upper bound~\eqref{eq:singleupbound} is subdivided
into two parts: First, in Section~\ref{sec:upboundgen}, we give
general conditions on a measure $M$ such that $M(\sigma_{X Y E})$ is
an upper bound on the rate $r_{\sigma_{X Y E}}$.  Second, in
Section~\ref{sec:concupbound}, we show that the measure $M$ defined by
the right hand side of~\eqref{eq:singleupbound} satisfies these
conditions.

\subsection{General properties of upper bounds} \label{sec:upboundgen}


Let $M$ be a real-valued function on the set of tripartite density
operators.  We show that $M(\sigma_{X Y E})$ is an upper bound on the
rate $r_{\sigma_{X Y E}}$ if the following conditions are satisfied.
(Here, we also write $M(X;Y;E)$ instead of $M(\sigma_{X Y E})$.
Moreover, if a random variable $X'$ is computed from $X$, we write $X'
\leftarrow X$.)
\begin{enumerate}
\item $M(\sigma_{X Y E}^{\otimes n}) \leq n M(\sigma_{X Y E})$, for
  any $n \in \bbN$.
    \label{prop:addone}
  \item $M(X';Y;E) \leq M(X;Y;E)$ for $X' \leftarrow X$.
    \label{prop:locA}
  \item $M(X;Y';E) \leq M(X;Y;E)$ for $Y' \leftarrow Y$.
    \label{prop:locB}
  \item $M(X C; Y C; E C) \leq M(X;Y;E)$ for $C \leftarrow X$.
    \label{prop:comA}
  \item There exists a function $\alpha$ with $\lim_{\eps \to 0}
    \alpha(\eps) = 0$ such that, for any state $\rho_{\bS_A \bS_B E}$
    describing an $\eps$-secure key pair of length $\ell$
    (cf.~\eqref{eq:epssecure}),
    \[
    M(\rho_{\bS_A \bS_B E}) \geq \bigl(1- \alpha(\eps) \bigr) \ell \ .
    \]
    \label{prop:norm}
\end{enumerate}

Consider an arbitrary secret-key agreement protocol and assume that
the protocol starts with $n$ copies of the state $\sigma_{X Y E}$.
Let $\rho^n_{\bS_A \bS_B E'}$ be the overall state of Alice's and
Bob's key $\bS_A$ and $\bS_B$, respectively, together with the
adversary's information $E'$ after the protocol execution. Then, using
properties~\ref{prop:addone}--\ref{prop:comA}, we find
\begin{equation} \label{eq:protsteps}
  n M(\sigma_{X Y E})
\geq
  M(\sigma_{X Y E}^{\otimes n})
\geq
  M(\rho^n_{\bS_A \bS_B E'}) \ .
\end{equation}
For any $n \in \bbN$, the resulting state must be $\eps(n)$-close to a
state describing a secret key of length $\ell(n)$, for $\eps(n)
\rightarrow 0$ as $n$ approaches infinity. Hence,
from~\eqref{eq:protsteps} and property~\ref{prop:norm},
  \[
   M(\sigma_{X YE})
  \geq
   \lim_{n \to \infty} \frac{\ell(n)}{n}
  =
    r(\sigma_{X Y E}) \ ,
  \]
  which concludes the proof.

\subsection{A concrete expression for the upper bound} \label{sec:concupbound}

Let $M$ be the measure defined by the right hand side
of~\eqref{eq:singleupbound}, i.e., for any tripartite density operator
$\sigma_{X Y E}$, $M(\sigma_{X Y E}) := M(X;Y;E)$ is given by
\[
  M(X;Y;E)
:=
  \sup_{\sigma_U^x, \sigma_V^x}
    \bigl(S(U | V E) - S(U | V Y) \bigr) \ .
\]
The goal of this section is to show that this measure satisfies the
conditions of Section~\ref{sec:upboundgen}, which implies that
$M(\sigma_{X Y E})$ is an upper bound on the secret-key rate
$r(\sigma_{X Y E})$.

Let us start with property~\ref{prop:addone}. It suffices to show
that, for any state $\sigma_{X Y E X' Y' E'}:=\sigma_{X Y E} \otimes
\sigma_{X' Y' E'}$,
  \[
    M(X X'; Y Y'; E E') \leq M(X;Y;E) + M(X';Y';E') \ ,
  \]
  i.e.,
  \begin{multline*}
    \sup_{(\tilde{U},\tilde{V}) \leftarrow (X,X')}
      S(\tilde{U}|\tilde{V} E E') - S(\tilde{U}|\tilde{V} Y Y') \\
  \leq
    \sup_{(U,V) \leftarrow X} S(U|V E) - S(U|V Y) \\
    \quad + \sup_{(U',V') \leftarrow X'} S(U'|V' E') - S(U'|V' Y') \ .
  \end{multline*}
  where $(U,V) \leftarrow X$ (and likewise $(U', V') \leftarrow X'$
  and $(\tilde{U}, \tilde{V}) \leftarrow (X,X')$) means that the
  density operators $\sigma_U^x$ and $\sigma_V^x$ used for the
  definition of $\sigma_{U V E}$ and $\sigma_{U V Y}$
  (cf.~\eqref{eq:upboundsigma}) are computed from the classical random
  variable $X$. The left hand side of this expression can be upper
  bounded by
  \begin{multline*}
    \sup_{(\tilde{U},\tilde{V}) \leftarrow (X,X')}
      S(\tilde{U}|\tilde{V} E E') - S(\tilde{U}|\tilde{V} Y E') \\
  \quad +
    \sup_{(\tilde{U},\tilde{V}) \leftarrow (X,X')}
      S(\tilde{U}|\tilde{V} Y E') - S(\tilde{U}|\tilde{V} Y Y') \ .
  \end{multline*}
  It thus remains to be shown that for any $(\tilde{U},\tilde{V})
  \leftarrow (X,X')$ there exists $(U,V) \leftarrow X$ such that
  \begin{equation} \label{eq:compfirst}
       S(\tilde{U}|\tilde{V} E E') - S(\tilde{U}|\tilde{V} Y E')
  \leq
       S(U|V E) - S(U|V Y)
  \end{equation}
  and, similarly, for any $(\tilde{U},\tilde{V}) \leftarrow (X,X')$
  there exists $(U',V') \leftarrow X'$ such that
  \begin{equation} \label{eq:compsecond}
      S(\tilde{U}|\tilde{V} Y E') - S(\tilde{U}|\tilde{V} Y Y')
  \leq
      S(U'|V' E') - S(U'|V' Y') \ .
  \end{equation}
  Inequality~\eqref{eq:compfirst} follows from the observation that
  $(\tilde{U}, \tilde{V}, E') \leftarrow X \leftarrow (Y,E)$ is a
  Markov chain~\footnote{Let $\sigma_{A B Z}$ be a tripartite quantum
    state of the form $\sigma_{A B Z} = \sum_{z} P_Z(z) \, \sigma_{A
      B}^z \otimes P_{\ket{z}}$, where $\{\ket{z}\}$ is a family of
    orthonormal vectors. We say that $A \leftarrow Z \leftarrow B$ is
    a Markov chain if $\sigma_{A B Z} = \sum_{z} P_Z(z) \,
    \sigma_{A}^z \otimes \sigma_{B}^z \otimes P_{\ket{z}}$, i.e., the
    state in the subsystem $A$ is fully determined by the classical
    value $z$.}, that is, we can set $U:=\tilde{U}$ and
  $V:=(\tilde{V},E')$, in which case the left hand side and the right
  hand side of~\eqref{eq:compfirst} become identical.
  Inequality~\eqref{eq:compsecond} follows similarly from the fact
  that $(\tilde{U}, \tilde{V}, Y) \leftarrow X' \leftarrow (Y',E')$ is
  a Markov chain, i.e., we can set $U':=\tilde{U}$ and
  $V':=(\tilde{V},Y)$ to obtain equality.
  
  To prove property~\ref{prop:locA}, that is, for any $X' \leftarrow
  X$,
  \begin{multline*}
    \sup_{(U',V') \leftarrow X'} S(U'|V' E) - S(U'|V' Y) \\
  \leq
    \sup_{(U,V) \leftarrow X} S(U|V E) - S(U|V Y) \ ,
  \end{multline*}
  it suffices to show that if $(U',V') \leftarrow X' \leftarrow (X, Y,
  E)$ is a Markov chain then $(U',V') \leftarrow X \leftarrow (Y, E)$
  is a Markov chain. This is true since $X' \leftarrow X \leftarrow
  (Y,E)$ is a Markov chain.
  
  For property~\ref{prop:locB}, we need to show that, for any $Y'
  \leftarrow Y$,
  \begin{multline*}
    \sup_{(U,V) \leftarrow X} S(U|V E) - S(U|V Y') \\
  \leq
    \sup_{(U,V) \leftarrow X} S(U|V E) - S(U|V Y) \ .
  \end{multline*}
  This is however a direct consequence of the strong subadditivity,
  implying that
  \[
    S(U|V Y') \geq S(U|V Y' Y) = S(U|V Y) \ ,
  \]
  where the equality is a consequence of the fact that $Y' \leftarrow
  Y \leftarrow (U,V)$ is a Markov chain.

  To prove property~\ref{prop:comA}, i.e., for $C \leftarrow X$,
  \begin{multline*}
    \sup_{(U',V') \leftarrow (X,C)} S(U'|V' E C) - S(U'|V' Y C) \\
  \leq
    \sup_{(U,V) \leftarrow X} S(U|V E) - S(U|V Y) \ ,
  \end{multline*}
  note that $(U', V', C) \leftarrow X \leftarrow (Y,E)$ is a Markov
  chain. We can thus set $U:=U'$ and $V:=(V',C)$, in which case the
  left hand side and the right hand side of the above expression
  become equal.

  It remains to be shown that property~\ref{prop:norm} holds.  Let
  $\sigma_U^x := P_{\ket{x}}$ and let $\sigma_V^x$ be an arbitrary
  state independent of $x$. Then, from Lemma~\ref{lem:SABbound},
  \begin{multline*}
    M(\bS_A;\bS_B;E)
  \geq
    S(\bS_A|E) - S(\bS_A|\bS_B) \\
  \geq
    S(\bS_A) - \sqrt{2 \eps} \ell - 1/e - S(\bS_A|\bS_B) \ ,
  \end{multline*}
  where $M(\bS_A;\bS_B;E) := M(\rho_{\bS_A \bS_B
    E})$.  The assertion then follows from the fact that
  \[
    I(\bS_A;\bS_B) \geq \bigl((1-\eps - 2h(\eps))\bigr) \ell \ .
  \]

\section{Examples: The six-state, BB84, and B92 protocols} \label{sec:examples}

To compute expression~\eqref{eq:singlebound} for the secret-key rate,
we have to optimize over the choices of the channels $U \leftarrow X$
and $V \leftarrow U$ used for the classical processing. Clearly, every
choice of these channels gives a lower bound on the rate.
Surprisingly, for the QKD protocols considered below, a good choice is
to define $U$ as a noisy version of $X$, while $V$ is set to a
constant, i.e., it can be discarded. For the protocol, this means
that, before doing error correction, Alice should simply add some
noise to her measurement data. Intuitively, this puts Bob into a
better position than Eve, since the effect of this noise on the
correlation between Alice and Eve is worse than on those between Alice
and Bob.



\subsection{Six-state protocol}

The six-state protocol~\cite{BeGi99} uses three different encodings,
defined by the $z$-basis $\{\ket{0}_z, \ket{1}_z\}$, the $x$-basis
$\{\ket{0}_x, \ket{1}_x\} := \{1/\sqrt{2}(\ket{0}_z \pm \ket{1}_z)\}$,
and the $y$-basis $\{\ket{0}_y, \ket{1}_y\} := \{1/\sqrt{2}(\ket{0}_z
\pm i \ket{1}_z)\}$.  Alice and Bob measure the QBER for each of these
encodings. This gives three conditions on the diagonal entries
$\lambda_1, \ldots, \lambda_4$ (with respect to the Bell basis) of the
states $\sigma_{A B}$ contained in the set~$\Gamma_\QBER$ over which
we have to minimize (see equation~\eqref{eq:singlebound}).  In
particular, if the QBER equals $Q$ for all encodings, we get
$\lambda_3 + \lambda_4 = Q$, $\lambda_2 + \lambda_4 = Q$, and
$\lambda_2 + \lambda_3 = Q$.  Together with the normalization, we
immediately find $\lambda_1 = 1-\frac{3}{2} Q$ and $\lambda_2 =
\lambda_3 = \lambda_4 = \frac{1}{2} Q$.

In order to evaluate the entropies occurring in
expression~\eqref{eq:singlebound}, we need to consider a purification
$\ket{\psi}_{A B E}$ of the diagonalization $\cD_2(\sigma_{A B})$ of
$\sigma_{A B}$, i.e.,
\[
  \ket{\psi}_{A B E} 
:= 
  \sum_{i=1}^4 
  \sqrt{\lambda_i} \ket{\Phi_i}_{A B} \otimes \ket{\nu_i}_E \ ,
\]
where $\ket{\Phi_1}_{A B}, \ldots, \ket{\Phi_4}_{A B}$ denote the Bell
states in Alice and Bob's joint system (with respect to the
$z$-basis~\footnote{We assume here that the encoding with respect to
  the $z$-basis is chosen with probability almost one (see also the
  discussion in Section~\ref{sec:lowbound} and~\cite{LoChAr00}) such
  that the number of bit pairs discarded in the sifting step is
  negligible.})  and where $\ket{\nu_1}_E, \ldots, \ket{\nu_4}_E$ are
some mutually orthogonal states in Eve's system. It is easy to verify
that, if Alice and Bob apply their measurements (with respect to the
$z$-basis), resulting in outcomes $x$ and $y$, respectively, the state
of Eve's system is given by $\ket{\theta^{x,y}}$, where
\begin{align*}
  \ket{\theta^{0,0}} 
& = 
  \frac{1}{\sqrt{2}} \bigl( \sqrt{\lambda_1} \ket{\nu_1}_E 
  + \sqrt{\lambda_2} \ket{\nu_2}_E \bigr) 
\\ 
 \ket{\theta^{1,1}} 
& = 
  \frac{1}{\sqrt{2}} \bigl( \sqrt{\lambda_1} \ket{\nu_1}_E
  - \sqrt{\lambda_2} \ket{\nu_2}_E \bigr) 
\\ 
 \ket{\theta^{0,1}} 
& = 
  \frac{1}{\sqrt{2}} \bigl( \sqrt{\lambda_3} \ket{\nu_3}_E 
  + \sqrt{\lambda_4} \ket{\nu_4}_E \bigr) 
\\ 
 \ket{\theta^{1,0}} 
& = 
  \frac{1}{\sqrt{2}} \bigl( \sqrt{\lambda_3} \ket{\nu_3}_E
  - \sqrt{\lambda_4} \ket{\nu_4}_E \bigr) \ .
\end{align*}
In particular, the density operators $\sigma_E^0$ and $\sigma_E^1$
describing Eve's system, if Alice has the value $0$ or $1$,
respectively, are given by $\sigma_E^0 = \frac{1}{2}
P_{\ket{\theta^{0,0}}} + \frac{1}{2} P_{\ket{\theta^{0,1}}}$ and
$\sigma_E^1 = \frac{1}{2} P_{\ket{\theta^{1,0}}} + \frac{1}{2}
P_{\ket{\theta^{1,1}}}$. We can write these states with respect to the
basis $\{\ket{\nu_0}_E, \ldots, \ket{\nu_3}_E\}$,
\[
  \sigma_E^x
=
  \left(\begin{matrix}
    \lambda_1 & \pm \sqrt{\lambda_1 \lambda_2} & 0 & 0 \\ 
    \pm \sqrt{\lambda_1 \lambda_2} & \lambda_2 & 0 & 0 \\
    0 & 0 & \lambda_3 & \pm \sqrt{\lambda_3 \lambda_4} \\
    0 & 0 & \pm \sqrt{\lambda_3 \lambda_4} & \lambda_4
  \end{matrix}\right)
\]
where $\pm$ is a plus sign if $x=0$ and a minus sign if $x=1$.

As mentioned above, we define $U$ as a noisy version of $X$, with
bit-flip probability $q$, i.e., $P_{U|X=0}(1) = P_{U|X=1}(0) = q$.
Moreover, $V$ is set to a constant, which means that it can simply be
omitted.  Using the fact that $S(U E) = H(U) + S(E|U)$, and,
similarly, $H(U Y) = H(U) + H(Y|U)$, the entropy difference on the
right hand in the supremum of~\eqref{eq:singlebound} is given by
\[
  S(U|E) - H(U|Y) = S(E|U) - S(E) - (H(Y|U) - H(Y))
\]
with
\begin{align*}
  S(E|U) 
& = 
    \textstyle \frac{1}{2} S\bigl((1-q) \sigma^0_E + q \sigma^1_E\bigr)
    + \frac{1}{2} S\bigl(q \sigma^0_E + (1-q) \sigma^1_E\bigr)
 \\
  S(E) 
& =
  \textstyle S\bigl(\frac{1}{2} \sigma^0_E + \frac{1}{2} \sigma^1_E\bigr) \ .
\end{align*}
Furthermore, $H(Y) = 1$ and 
\[
  H(Y|U) = h[ q (1-Q) + (1-q) Q] \ ,
\]
where $h$ is the binary entropy function.

These expressions can easily be evaluated numerically. For an optimal
choice of the parameter $q$, we get a positive secret-key rate if $Q
\leq 0.141$.  Without the pre-processing, we obtain the known bound $Q
\leq 0.126$~\cite{Lo01} (see Fig.~\ref{fig:comp}). Remarkably, this
bound has already been improved to $Q \leq 0.127$~\cite{Lo01} using
degenerate quantum codes, which can be interpreted as a certain type
of pre-processing.

Another method to obtain conditions on the set $\Gamma_\QBER$
in~\eqref{eq:singlebound} is to use some additional symmetrization.
For this, we consider the operator $\cD_1$ as defined by~\eqref{D1}
with $A_1=V_x, A_2=V_y, A_3=V_z$ and $B_1=V_x,B_2=V_y^\dagger,
B_3=V_z$, where $V_x,V_y,V_z$ denote the unitary operators
transforming the $z$-basis into the $x,y,z$-basis, respectively.  This
implies that $\cD_2(\cD_1(\sigma_{A B})) =\lambda_1
P_{\ket{\Phi^+}}+\lambda_2 P_{\ket{\Phi^-}}+\lambda_3
P_{\ket{\Psi^+}}+\lambda_4 P_{\ket{\Psi^-}}$, where
$\lambda_3+\lambda_4=2\lambda_2$. As explained in~\cite{KrGiRe04}, we
can, instead of $\cD_2$, apply another symmetrization operation ${\cal
  D}^\prime_2(\rho)$, e.g., \bea {\cal D}^\prime_2(\rho)=\sum_{l}
O^\prime_l \otimes O^\prime_l \, \rho \, (O^\prime_l)^\dagger \otimes
(O^\prime_l)^\dagger, \eea where $O^\prime_{l}\in \{UV : U \in \{\one,
\sigma_z, \allowbreak \mbox{diag}(-i,1), \allowbreak
\mbox{diag}(i,1)\} \allowbreak \, \text{and} \, V\in
\{\one,\sigma_x\}\}$. Apart from depolarizing any state to a
Bell-diagonal state, this map also equalizes the coefficients
$\lambda_3$ and $\lambda_4$ in~\eqref{rho1}. This implies that
$\cD'_2(\cD_1(\Gamma_\QBER) = \{(1-3Q/2)P_{\ket{\Phi^+}}+Q/2(
P_{\ket{\Phi^-}}+P_{\ket{\Psi^+}}+P_{\ket{\Psi^-}})\}$. Thus, using
this method, we find right away all the necessary conditions on the
set $\Gamma_\QBER$.

Finally, we can use~\eqref{eq:upboundcl} to compute an upper bound on
the secret-key rate of the one-way six-state protocol. Let again
$\ket{\theta^{0,0}}$ and $\ket{\theta^{1,1}}$ be the states of Eve
conditioned on the event that Alice and Bob have the values $(0,0)$
and $(1,1)$, respectively. If the adversary applies a von Neumann
measurement with respect to projectors along $\frac{1}{\sqrt{2}}
(\ket{\theta^{0,0}} + \ket{\theta^{1,1}})$ and $\frac{1}{\sqrt{2}}
(\ket{\theta^{0,0}} - \ket{\theta^{1,1}})$, resulting in $Z$, we get
$r(X,Y,Z) = 0$ whenever $Q \geq 0.163$.

\setlength{\unitlength}{0.4mm}

\newcommand*{\markd}[2]{
  \put(#1,-2){\line(0,1){4}}
  \put(#1,-8){\makebox(0,0)[c]{#2}}
}

\newcommand*{\marku}[2]{
  \put(#1,-20){\line(0,1){22}}
  \put(#1,-26){\makebox(0,0)[c]{#2}}
}

\newcommand*{\markdd}[2]{
  \put(#1,-8){\line(0,1){10}}
  \put(#1,-14){\makebox(0,0)[c]{#2}}
}

\newcommand*{\markubreak}[2]{
  \put(#1,-20){\line(0,1){2}}
  \put(#1,-10){\line(0,1){12}}
  \put(#1,-26){\makebox(0,0)[c]{#2}}
}

\newcommand*{\markddd}[2]{
  \put(#1,-32){\line(0,1){34}}
  \put(#1,-38){\makebox(0,0)[c]{#2}}
}

\begin{figure} 
\begin{picture}(200,70)(0, -35)
  \put(0,0){\line(1,0){24}}
  \put(28,0){\vector(1,0){177}}
  \put(200,-8){\makebox(0,0)[c]{$Q$}}

  \put(10,-14){\makebox(0,0)[l]{new bounds}}
  \put(10,-26){\makebox(0,0)[l]{previous bounds}}
  \put(10,-38){\makebox(0,0)[l]{C.K.}}

  \markd{0}{0}
  \markdd{122}{14.1}
  \markdd{164}{16.3}

  \marku{92}{12.7}
  \markubreak{172}{16.6}

  \markddd{154}{15.7}

  \put(0,4){\makebox(128.2,20)[cc]{$r>0$}}
  \put(0,4){\line(1,0){24}}
  \put(0,24){\line(1,0){24}}

  \put(28,4){\line(1,0){94}}  
  \put(28,24){\line(1,0){94}}

  \put(0,4){\line(0,1){20}}
  \put(122,4){\line(0,1){20}}

  \put(164.3,4){\makebox(48,20)[cc]{$r=0$}}   
  \put(164.3,4){\line(1,0){38}}
  \put(164.3,24){\line(1,0){38}}
  \put(164.3,4){\line(0,1){20}}

  \put(122,4){\makebox(42,20)[cc]{$?$}} 
  \put(122,4){\dashbox(42,0)[cc]{}}
  \put(122,24){\dashbox(42,0)[cc]{}}

\end{picture}

\caption{\label{fig:comp} Lower and upper bounds on the maximally tolerable QBER
  $Q$ in percent for the six-state protocol. The last line (C.K.)
  indicates the QBER such that $I(X;Y) = I(X;Z) = I(Y;Z)$ where $X$
  and $Y$ is Alice's and Bob's classical information, respectively,
  and where $Z$ is the classical information that Eve can gain in an
  individual attack.}

\end{figure}
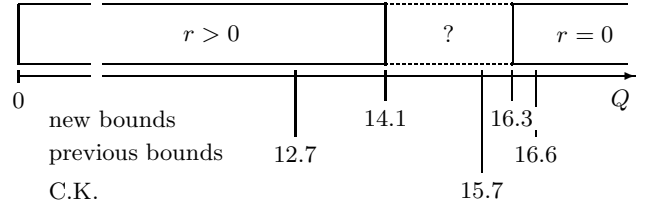

\subsection{BB84}

The BB84 protocol~\cite{BB84} is very similar to the six-state
protocol, but uses only two of the three bases for the encoding.
Hence, one only gets two conditions on the diagonal entries
$\lambda_1, \ldots, \lambda_4$ (with respect to the Bell basis) of the
density operator $\sigma_{A B}$, namely $\lambda_3 + \lambda_4 = Q$
and $\lambda_2 + \lambda_4 = Q$. Hence, the set $\Gamma_{\QBER}$
contains all states with diagonal entries $\lambda_1 = 1-2 Q +
\lambda_4$ and $\lambda_2 = \lambda_3 = Q - \lambda_4$, for any
$\lambda_4 \in [0,Q]$.

\begin{figure}[t] 
  \centering
  \includegraphics[width=\linewidth]{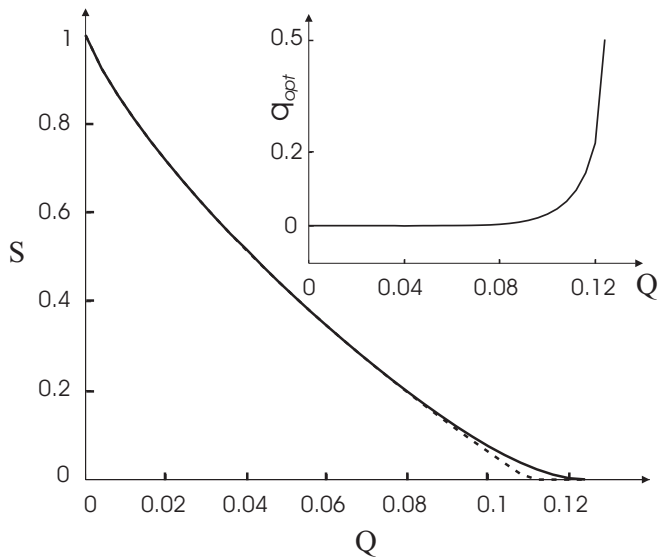}
  \caption{\label{fig:BB84}
    Lower bound on the secret-key rate of the BB84 protocol as a
    function of the QBER $Q$. The dashed line represents the known
    result \cite{BB84}, whereas the full line shows our new lower
    bound. The insert shows the optimal value $q_\mathrm{opt}$ for the
    probability by which Alice has to flip her bits in the
    pre-processing phase.}
\end{figure}

The evaluation of~\eqref{eq:singlebound} now follows the same lines as
described above for the six-state protocol.  A straightforward
calculation shows that, independently of the amount of noise added in
the pre-processing, expression~\eqref{eq:singlebound} takes its
minimum for $\lambda_4 = Q^2$.  When optimizing over the preprocessing
(i.e., the amount of noise introduced by Alice) we get a positive rate
if $Q \leq 0.124$ (see Fig.~\ref{fig:BB84}).  Note that, without the
pre-processing, we obtain $Q \leq 0.110$, which is exactly the bound
due to Shor and Preskill~\cite{ShPr00}.  Computing the upper
bound~\eqref{eq:upboundcl} reproduces the known result saying that the
(one-way) secret-key rate is zero if $Q \geq 0.146$~\cite{FuGr97}.

\subsection{B92}

In contrast to the BB84 and the six-state protocol, Alice uses two
non-orthogonal states~\cite{Be92}
$\ket{\varphi^0}=\alpha\ket{0}+\beta\ket{1}$ and
$\ket{\varphi^1}=\alpha\ket{0}-\beta\ket{1}$ to encode her bit-values
$0$ and $1$, respectively, where $\alpha$ and $\beta$ are (without
loss of generality) real coefficients with $\alpha^2 + \beta^2 = 1$.
Bob randomly applies a measurement with respect to the basis
$\{\ket{\varphi^0}, \ket{\varphi^0}^\lperp\}$ or $\{\ket{\varphi^1},
\ket{\varphi^1}^\lperp\}$, where $\ket{\varphi^x}^\lperp$ denotes the
normalized vector orthogonal to $\ket{\varphi^x}$, for $x=0,1$. He
then assigns the bit values $0$ and $1$ to the measurement outcomes
$\ket{\varphi^1}^\lperp$ and $\ket{\varphi^0}^\lperp$, respectively.
In the sifting step, Alice and Bob discard all bit pairs where Bob
measured $\ket{\varphi_0}$ or $\ket{\varphi_1}$.

In order to evaluate expression~\eqref{eq:singlebound}, we will rely
on some of the calculations presented in~\cite{ChReEk04}.  We first
need a description of Alice and Bob's data bits after the sifting
step. Note that, in contrast to the BB84 or the six-state protocol,
the sifting only depends on the measurement outcomes of Bob.
Therefore, we consider the state obtained from the operation $\one_A
\otimes B$, with $B := \ket{0} \bra{\varphi^1}^\lperp + \ket{1}
\bra{\varphi^0}^\lperp$, applied to each of the qubit pairs. Note that
this corresponds to the application of the map $\cD_1$
(see~\eqref{D1}). $\Gamma_\QBER$ is then defined as the set of all
states $\sigma_{A B}$ which can result from this operation (applied to
any two-qubit density operator which corresponds to a collective
attack of Eve) and, in addition, are compatible with the QBER.
In~\cite{ChReEk04}, explicit conditions on the diagonal entries (with
respect to the Bell basis) of these states have been computed.  In
particular, the first two diagonal entries are $\lambda_1 = (1-Q)
\frac{1+s}{2}$ and $\lambda_2 = (1-Q) \frac{1-s}{2}$ where $s$ is the
scalar product between the states of the adversary, conditioned on the
event that Alice and Bob have the values $(0,0)$ and $(1,1)$,
respectively. This characterization is already sufficient to obtain
reasonable lower bounds on the rate~\eqref{eq:singlebound}.

\begin{figure}[t]
  \centering \includegraphics[width=\linewidth]{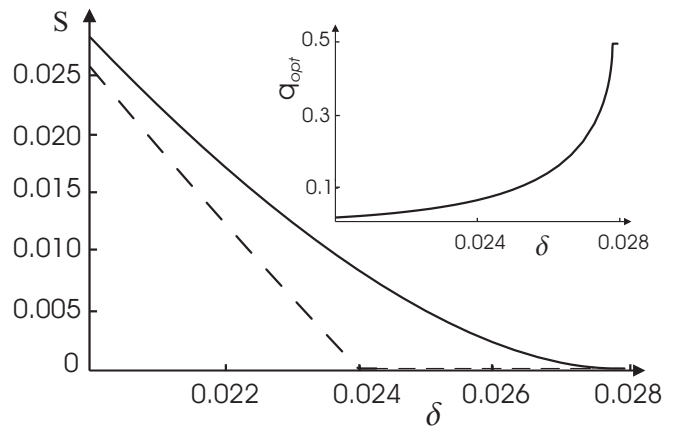}
  \caption{\label{fig:B92}
    Lower bound on the secret-key rate of the B92 protocol, for
    $\alpha = 0.38$ (see text for an explanation of the
    parameter~$\delta$).  The dashed line represents the known result
    without pre-processing~\cite{ChReEk04}, whereas the solid line is
    our new lower bound on the rate when Alice additionally adds noise
    $q_{\mathrm{opt}}$ to her measurement data.}
\end{figure}

Similarly to the previous examples, adding noise on Alice's side turns
out to be useful.  The results of our computations are summarized in
Fig.~\ref{fig:B92}, parameterized by the noise $\delta$ of a
corresponding depolarizing channel $\rho \mapsto (1-2 \delta) \rho +
\delta \one$ \footnote{For any given value of the QBER, the value
  $\delta$ is defined as the parameter of a depolarizing channel $\rho
  \mapsto (1-2 \delta) \rho + \delta \one$ which produces the same
  QBER when employing the protocol.}. The rate is positive as long as
$\delta \leq 0.0278$ (compared to $\delta \lessapprox 0.0240$ without
noise~\cite{ChReEk04,TaKoIm03}).  Within the region shown in the
figure, the relation between the parameter $\delta$ and the QBER is
$Q\approx 2 \delta$ \footnote{In general we have $Q= \delta /
  (\gamma^2 (1-2 \delta) + 2 \delta)$, where $\gamma^2 = 4 \alpha^2
  (1-\alpha^2)$.}.

\section{Conclusions and open problems}

We have analyzed a general class of QKD protocols with one-way
classical post-processing, thereby using a technique which is not
based on entanglement purification. We have shown that, in order to
guarantee security against the most general attacks, it is sufficient
to consider collective attacks. Moreover, we have derived a new
general lower bound on the secret-key rate
(formula~\eqref{eq:singlebound}) which is very similar to the
well-known expression for the classical one-way secret-key rate due to
Csisz\'ar and K\"orner~\cite{CsiKoe78}.  While the latter applies if
the information of the adversary is purely classical (i.e., if she is
restricted to individual attacks), expression~\eqref{eq:singlebound}
can be seen as a quantum version of it.

In order to evaluate~\eqref{eq:singlebound}, one only needs to
optimize over a certain set of two-qubit density operators, which is
characterized by the possible collective attacks on the specific
protocol.  We have illustrated this for some of the most popular QKD
schemes, namely the BB84, the six-state, and the B92-protocol, with
one-way classical post-processing, say, from Alice to Bob.
Surprisingly, our results imply that the performance of these
protocols can be increased if Alice introduces noise to her
measurement data. In particular, we get new lower bounds on the
maximum tolerated channel noise which are between 10 and 15 percent
larger than the previously known ones.

While our method allows to exactly analyze the security of a general
class of QKD protocols with one-way post-processing, it is still an
open problem to identify the protocols which achieve the maximum rate.
In particular, we do not know whether a bit-wise pre-processing is
optimal, or whether it might be more advantageous for Alice and Bob to
process larger blocks. Note, however, that the upper
bound~\eqref{eq:singleupbound} on the secret-key rate of one-way
protocols essentially has the same form as the lower
bound~\eqref{eq:singlebound}, but involves a maximization over certain
quantum states instead of only classical random variables. The
question of whether bit-wise pre-processing is optimal thus reduces to
the problem of proving that these two expressions are equal.

\section{Acknowledgment}

N.~G.\ and B.~K.\ thank the Swiss NCCR, ``Quantum photonics'', and the
European IST project SECOQC. We would also like to thank Robert
K\"onig and Valerio Scarani for many helpful comments.

\appendix

\section{Smooth R\'enyi entropy} \label{sec:smooth}

\subsection{Basic properties}

Smooth R\'enyi entropy has been introduced in~\cite{RenWol04} in order
to characterize fundamental properties of classical random variables.
For instance, the $\eps$-smooth R\'enyi entropy of order $0$ of a
random variable $X$ conditioned on $Y$, denoted $H_0^\eps(X|Y)$,
measures the minimum length of an encoding $C$ of $X$ such that $X$
can be reconstructed from $C$ and $Y$, except with probability roughly
$\eps$.  Similarly, the $\eps$-smooth R\'enyi entropy of order $2$,
denoted $H_2^\eps(X|Y)$, quantifies the amount of uniform randomness
independent of $Y$ that can be extracted from $X$ (with probability
roughly $1-\eps$).

In~\cite{RenKoe05}, the notion of smooth R\'enyi entropy has been
generalized to quantum states. For a density operator $\rho$, we
denote by $S_\alpha^\eps(\rho)$, the $\eps$-smooth R\'enyi entropy of
order $\alpha$ of $\rho$. Similar to the von Neumann entropy,
$S_\alpha^\eps(\rho)$ is defined as the (classical) smooth R\'enyi
entropy of the eigenvalues of $\rho$, interpreted as a probability
distribution.  We also write $S_\alpha^\eps(U V)$ instead of
$S_\alpha^\eps(\rho_{U V})$ and, similarly, $S_\alpha^\eps(U)$ instead
of $S_\alpha^\eps(\rho_{U})$, where $\rho_{U}$ is the partial state
$\rho_{U} := \tr_{V}(\rho_{U V})$.

We start reviewing some basic properties of smooth R\'enyi entropy of
quantum states. The proofs can be found in~\cite{RenWol04}
and~\cite{RenKoe05}.  Most of these properties are very analogous to
the properties of the von Neumann entropy $S(\cdot)$. For instance, if
$\rho_{U V}$ is a state on $\cH_U \otimes \cH_V$, then the difference
between $S_\alpha^\eps(U V)$ and $S_\alpha^\eps(U)$ is bounded by the
entropy of $V$, which corresponds to the well known fact that $S(U) -
S(V) \leq S(U V) \leq S(U) + S(V)$: For $\alpha=2$, we have
\begin{align} \label{eq:SABinfupbound}
  S_2^\eps(U V) & \leq \Hq^{\eps+\eps'}_2(U) + S_0^{\eps'}(V)
\\ \label{eq:SABinflowbound}
  S_2^{\eps+\eps'}(U V) & \geq S_2^\eps(U) - S_0^{\eps'}(V)
\\ \intertext{and, similarly, for $\alpha=0$,} \label{eq:SABzeroupbound}
  S_0^{\eps+\eps'}(U V) & \leq S_0^\eps(U) + S_0^{\eps'}(V)
\\ \label{eq:SABzerolowbound}
  S_0^{\eps}(U V) & \geq S^{\eps+\eps'}_0(U) - S_0^{\eps'}(V) \ .
\end{align}
Consider now a bipartite state $\rho_{U Z}$ on $\cH_U \otimes \cH_Z$
where the second part is purely classical, i.e.,
\[
  \rho_{U Z} = \sum_{z} P_Z(z) \, \rho^z_U \otimes P_{\ket{z}} \ ,
\]
for some probability distribution $P_Z$ and a family of orthonormal
vectors $\{\ket{z}\}_{z}$ on $\cH_Z$. Then, the smooth R\'enyi entropy
cannot increase when conditioning on $Z$, that is,
\begin{equation} \label{eq:Hzerotrbound}
  S^\eps_\alpha(U|Z) \leq S_\alpha^\eps(U) \ ,
\end{equation}
for $\alpha = 0$ and $\alpha=2$. The following inequalities can be
interpreted as extensions of the chain rule $S(U Z) = S(U|Z) + S(Z)$
to smooth R\'enyi entropy:
\begin{align} \label{eq:Hinfcondupbound}
  S_2^\eps(U|Z)
    & \leq S_2^{\eps+\eps'}(U Z) - H_2^{\eps'}(Z)
\\  \label{eq:Hinfcondlowbound}
  S_2^{\eps+\eps'+\eps''}(U|Z)
    & > S_2^{\eps'}(U Z) - H_0^{\eps''}(Z) - 2 \log(1/\eps)
\\ 
\label{eq:Hzerocondlowbound}
  S_0^\eps(U|Z) & \geq S_0^{\eps+\eps'}(U Z) - H_0^{\eps'}(Z)
\\
  S_0^{\eps+\eps'+\eps''}(U|Z)
    & < S_0^{\eps'}(U Z) - H_2^{\eps''}(Z) + 2 \log(1/\eps) \ .
\end{align}
More generally, let $\rho_{U Z V}$ be a density operator on $\cH_U
\otimes \cH_Z \otimes \cH_V$ such that the states on $\cH_U$ and
$\cH_V$ only depend on the classical subsystem $\cH_Z$, i.e., there
exist density operators $\rho^z_U$ and $\rho^z_V$ on $\cH_U$ and
$\cH_V$, respectively, such that
\[
  \rho_{U V Z}
=
  \sum_{z \in \cZ}
    P_Z(z) \rho^z_U \otimes \rho^z_V \otimes P_{\ket{z}} \ ,
\]
where $P_Z$ is a probability distribution and $\{\ket{z}\}_{z \in
  \cZ}$ a family of orthonormal vectors on $\cH_Z$. Then
\begin{align} \label{eq:markov}
  S_2^{\eps + \eps' }(U V Z)
& \geq
  S_2^{\eps}(U|Z) + S_2^{\eps'}(V Z)
\\
  S_0^{\eps + \eps' }(U V Z)
& \leq
  S_0^{\eps}(U|Z) + S_0^{\eps'}(V Z) \ .
\end{align}

The following identities are useful to determine the conditional
smooth R\'enyi entropy $S_\alpha^\eps(U|Z)$ if the smooth R\'enyi
entropy $S_\alpha^\eps(U|Z=z)$, conditioned on certain values $z$, is
known. For any $z \in \cZ$, let $\eps_z := \eps \cdot P_Z(z)$.  Then
\begin{align} \label{eq:Hinfcondvalue}
  S_2^{\eps_z}(U|Z) & \leq \Hq^\eps_2(U|Z=z)
\\ \label{eq:Hzerocondvalue}
  S_0^{\eps_z}(U|Z) & \geq S_0^\eps(U|Z=z) \ .
\end{align}
Additionally, for any set $\bar{\cZ} \subset \cZ$ such that $\Pr_{z}[z
\in \bar{\cZ}] \geq 1-\eps$,
\begin{align} \label{eq:Hinfeventbound}
  S_2^{\eps+\eps'}(U|Z)
& \geq
  \min_{z \in \bar{\cZ}} S_2^{\eps'}(U|Z=z)
\\ \label{eq:Hzeroeventbound}
  S_0^{\eps+\eps'}(U|Z)
& \leq
  \max_{z \in \bar{\cZ}} S_0^{\eps'}(U|Z=z) \ .
\end{align}

Similarly to the von Neumann entropy, the smooth R\'enyi entropy can
only increase when applying a unital quantum operation
$\cE$~\footnote{A quantum operation $\cE$ is \emph{unital} if $\cE$ is
  trace-preserving and if the fully mixed state is a fixed point of
  $\cE$. Formally, if $\rho \mapsto \sum_{z} E_z \rho E_z^\dagger$ is
  the operator-sum representation of $\cE$, then $\sum_{z} E_z^\dagger
  E_z = E_z E_z^\dagger = \one$.}, that is,
\begin{equation} \label{eq:Hmeasinc}
  S_\alpha^\eps(\cE(\rho_U)) \geq \Hq_\alpha^\eps(\rho_U) \ .
\end{equation}
for any $\alpha \in \bbR^+$ and $\eps \in \bbR^+$.

The smooth R\'enyi entropies of order $\alpha$ are related for
different values of $\alpha$. In particular, we have
\begin{equation} \label{eq:Salphabeta}
  S_2^\eps(U) \lessapprox S_0^\eps(U) \ ,
\end{equation}
where the approximation holds up to $O(\eps)$. Finally, the smooth
R\'enyi entropy of an $n$-fold product state $\rho^{\otimes n}$
approaches the von Neumann entropy. Formally, for any $\alpha \in
\bbR^+$ and $\eps \in \bbR^+$,
\begin{equation} \label{eq:typsec}
    \bigl|\Hq_\alpha^{\eps}(\rho^{\otimes n}) - n \Hq(\rho)\bigr|
  \leq
    O(\log(1/\eps)) \ .
\end{equation}

\subsection{Smooth R\'enyi entropy and measurements}

Let $\cE$ be a measurement defined by a family of operators
$\{E_z\}_{z \in \cZ}$. Let $\rho_{\tilde{U}}:=\cE(\rho_U) = \sum_z E_z
\rho_U E_z^\dagger$ be the state of the quantum system after applying
$\cE$ to a density operator $\rho_{U}$, and let $Z$ be the classical
measurement outcome, i.e., $P_Z(z) := \tr(E_z \rho_U E_z^\dagger)$,
for $z \in \cZ$. We have seen in the previous section
(see~\eqref{eq:Hmeasinc}) that the entropy $S_\alpha^\eps(\tilde{U})$
of $\rho_{\tilde{U}}$ can only be larger than the entropy
$S_\alpha^\eps(U)$ of $\rho_{U}$ if $\cE$ is unital.  The following
lemma states that the maximum increase of the smooth R\'enyi entropy
when applying $\cE$ is bounded by the entropy $H_0^\eps(Z)$ of the
classical measurement outcome $Z$.

\begin{lemma} \label{lem:measbound}
  Let $\rho_{\tilde{U}}$ be the state obtained when applying the
  trace-preserving measurement $\cE$ to $\rho_{U}$ and let $Z$ be the
  classical outcome. Then, for $\eps, \eps' \in \bbR^+$,
 \begin{align} \label{eq:Hinfmeasdec}
    S_2^\eps(\tilde{U})
  & \leq
    S_2^{\eps+\eps'}(U) + H_0^{\eps'}(Z)
  \\ 
\label{eq:Hzeromeasdec}
    S_0^{\eps+\eps'}(\tilde{U})
  & \leq
    S_0^{\eps}(U) + H_0^{\eps'}(Z) \ .
  \end{align}
\end{lemma}

\begin{proof}
  Let $T$ be the linear operation from $\cH_U$ to $\cH_{\tilde{U}}
  \otimes \cH_Z$ defined by
  \[
    T : \quad \ket{\varphi}
  \longmapsto
    \sum_{z \in \cZ} (E_z \ket{\varphi}) \otimes \ket{z} \ ,
  \]
  for any $\ket{\varphi} \in \cH_U$, where $\{\ket{z}\}_z$ is a family
  of orthonormal vectors in $\cH_Z$. Let $\rho'_{\tilde{U} Z} := T
  \rho_{U} T^\dagger$. It is easy to verify that $\rho_{\tilde{U}} =
  \tr_Z(\rho'_{\tilde{U} Z})$, and that the eigenvalues of $\rho'_{Z}$
  correspond to the probabilities $P_Z(z)$. Hence, since the smooth
  R\'enyi entropy of quantum states is defined by the classical smooth
  R\'enyi entropy of its eigenvalues, we have
  $S_\alpha^{\eps'}(\rho'_Z) = H_\alpha^{\eps'}(Z)$.  Moreover,
  because $\cE$ is trace-preserving, i.e., $\sum_{z \in \cZ}
  E_z^\dagger E_z = \one_{U}$, we have $T^\dagger T = \one_{U}$.
  Consequently, $\rho'_{\tilde{U} Z}$ has the same eigenvalues as
  $\rho_{U}$, i.e., $S_\alpha^\eps(\rho'_{\tilde{U} Z}) =
  S_\alpha^\eps(\rho_{U})$.  Hence, using~\eqref{eq:SABinflowbound},
  we find
  \[
  \begin{split}
    S_2^\eps(\rho_{\tilde{U}})
  =
    S_2^\eps(\tr_Z(\rho'_{\tilde{U} Z}))
  & \leq
    S_2^{\eps+\eps'}(\rho'_{\tilde{U} Z}) + S_0^{\eps'}(\rho'_{Z}) \\
  & =
    S_2^{\eps+\eps'}(\rho_{U}) + H^{\eps'}_0(Z) \ ,
  \end{split}
  \]
  which concludes the proof of~\eqref{eq:Hinfmeasdec}.
  Inequality~\eqref{eq:Hzeromeasdec} follows by the same argument,
  where~\eqref{eq:SABinflowbound} is replaced
  by~\eqref{eq:SABzerolowbound}.
\end{proof}

A similar relation holds between the smooth R\'enyi entropy
$S_\alpha^\eps(U)$ of the original quantum state $\rho_U$ and the
entropy $S_\alpha^\eps(\tilde{U}|Z)$ of the state $\rho_{\tilde{U}}$
after the measurement, conditioned on the classical outcome $Z$.
Lemma~\ref{lem:projbound} below states that the difference between
these entropies is roughly bounded by the entropy $H_0^\eps(Z)$ of
$Z$.

\begin{lemma} \label{lem:projbound}
  Let $\rho_{\tilde{U}}$ be the state obtained when applying a von
  Neumann measurement $\cE$ to a state $\rho_{U}$. Let
  $S_\alpha^\eps(\tilde{U}|Z)$ be the entropy of $\rho_{\tilde{U}}$,
  conditioned on the classical outcome $Z$. Then, for $\eps, \eps',
  \eps'' \in \bbR^+$,
  \begin{align} \label{eq:Hinfmeaslowbound}
    S_2^{\eps+\eps'}(U)
  & \geq
    S_2^{\eps}(\tilde{U}|Z) - H_0^{\eps'}(Z)
  \\ \label{eq:Hinfmeasupbound}
    S_2^{\eps}(U)
  & <
    S_2^{\eps+\eps'+\eps''}(\tilde{U}|Z)
    + H_0^{\eps'}(Z) + 2 \log(1/\eps'')
  \\ \intertext{and}  \label{eq:Hzeromeasupbound}
    S_0^{\eps+\eps'}(U)
  & \leq
    S_0^\eps(\tilde{U}|Z) + H_0^{\eps'}(Z)
  \\ \label{eq:Hzeromeaslowbound}
    S_0^\eps(U)
  & \geq
    S_0^{\eps+\eps'}(\tilde{U}|Z) - H_0^{\eps'}(Z) \ .
  \end{align}
\end{lemma}

\begin{proof}
  Let $E_z$ be the projectors defined by the measurement $\cE$ and let
  $\rho_{\tilde{U} Z}$ be the state as defined in the proof of
  Lemma~\ref{lem:measbound}.  Since, by assumption, the ranges of the
  operators $E_z$, for $z \in \cZ$, are mutually orthogonal, the
  states $\rho_{\tilde{U} Z}$ and $\rho_{\tilde{U}}$ have the same
  eigenvalues and thus $S_\alpha^{\bar{\eps}}(\tilde{U} Z) =
  S_\alpha^{\bar{\eps}}(\tilde{U})$.  Using this identity,
  \eqref{eq:Hinfmeaslowbound} follows from~\eqref{eq:Hinfmeasdec}
  and~\eqref{eq:Hzerotrbound},
  \[
    S_2^{\eps+\eps'}(U)
  \geq
    S_2^{\eps}(\tilde{U})-H_0^{\eps'}(Z) \\
  \geq
    S_2^{\eps}(\tilde{U}|Z)-H_0^{\eps'}(Z) \ .
  \]
  Similarly, \eqref{eq:Hinfmeasupbound} follows
  from~\eqref{eq:Hmeasinc} and~\eqref{eq:Hinfcondlowbound},
  \begin{multline*}
    S_2^{\eps}(U)
  \leq
    S_2^{\eps}(\tilde{U})
  =
    S_2^{\eps}(\tilde{U} Z) \\
  <
    S_2^{\eps+\eps'+\eps''}(\tilde{U}|Z) + H_0^{\eps'}(Z)
    + 2 \log(1/\eps'') \ .
  \end{multline*}
  To prove~\eqref{eq:Hzeromeasupbound}, we use~\eqref{eq:Hmeasinc}
  and~\eqref{eq:Hzerocondlowbound},
  \begin{multline*}
    S_0^{\eps+\eps'}(U)
  \leq
    S_0^{\eps+\eps'}(\tilde{U})
  =
    S_0^{\eps+\eps'}(\tilde{U} Z) \\
  \leq
    S_0^{\eps}(\tilde{U}|Z) + H_0^{\eps'}(Z) \ .
  \end{multline*}
  Finally, \eqref{eq:Hzeromeaslowbound} follows
  from~\eqref{eq:Hzeromeasdec} and~\eqref{eq:Hzerotrbound},
  \[
    S_0^{\eps}(U)
  \geq
    S_0^{\eps+\eps'}(\tilde{U}) - H_0^{\eps'}(Z)
  \geq
    S_0^{\eps+\eps'}(\tilde{U}|Z) - H_0^{\eps'}(Z) \ .
  \]
\end{proof}

\subsection{The smooth R\'enyi entropy of symmetric states} \label{app:symstates}

The goal of this section is to derive an expression for the smooth
R\'enyi entropies of a symmetric state over $n$ subsystems in terms of
the von Neumann entropy of a corresponding state over only \emph{one}
subsystem.

Let $\sigma_1, \ldots, \sigma_d$ be density operators on $\cH_U$ and
let $\rho_{\bU}^n$ be the symmetric state over $\cH_U^{\otimes n}$
defined by
\begin{equation} \label{eq:rhoU}
  \rho_{\bU}^n
:=
  \cP_n\bigl(\sum_{\bn \in \Gamma^n_{d}}
    \mu_{\bn} \,
      \sigma_1^{\otimes n_1} \otimes \cdots
        \otimes \sigma_d^{\otimes n_d}
  \bigr)
\end{equation}
where, for any $\bn \in \Gamma^n_d := \{(n_1, \ldots, n_d): \sum_i n_i
= n\}$, $\mu_{\bn}$ are nonnegative coefficients such that $\sum_{\bn}
\mu_{\bn} = 1$.

Similarly, for any $d$-tuple $\lambda = (\lambda_1, \ldots \lambda_d)$
over $\bbR^+$, let $\sigma_U[\lambda]$ be the density operator on
$\cH_U$ defined by
\begin{equation} \label{eq:sigmaU}
  \sigma_U[\lambda]
:=
  \sum_i \lambda_i \sigma_i \ .
\end{equation}

Let $\cE$ be a quantum operation from $\cH_U$ to $\cH_V$.  The
following lemma gives a relation between the smooth R\'enyi entropy of
the symmetric state obtained by applying $\cE$ to each of the
subsystems of a purification of $\rho_U^n$ and the von Neumann
entropy of the state obtained by applying $\cE$ to a purification of
$\sigma_U[\lambda]$.

\begin{lemma} \label{lem:Hqcalc}
  Let $\rho_{\bU \bW}^n$ be a purification of the state $\rho_\bU^n$
  defined by~\eqref{eq:rhoU} with coefficients $\mu_{\bn}$ and let
  $\rho_{\bV \bW}^n := {(\cE \otimes \one_{W})}^{\otimes n}(\rho_{\bU
    \bW}^n)$. Similarly, for any $d$-tuple $\lambda$, let $\sigma_{U
    W}[\lambda]$ be a purification of the state $\sigma_U[\lambda]$
  defined by~\eqref{eq:sigmaU} and let $\sigma_{V W}[\lambda] := {(\cE
    \otimes \one_{W})} (\sigma_{U W}[\lambda])$.  Let $\bar{\Gamma}$
  be a subset of $\Gamma^n_d$ such that $ \sum_{\bn \in \bar{\Gamma}}
  \mu_{\bn} \geq 1-\frac{\eps}{2}$.  Then
  \begin{align*}
    S_2^\eps(\rho_{\bV \bW}^n)
  & \gtrapprox
    n \min_{\lambda}
      \Hq(\sigma_{\bV \bW}[\lambda])
  \\
    S_0^\eps(\rho_{\bV \bW}^n)
  & \lessapprox
    n \max_{\lambda}
      \Hq(\sigma_{\bV \bW}[\lambda])
  \end{align*}
  where the minimum and maximum are taken over all $\lambda =
  (\lambda_1, \ldots, \lambda_d)$ such that $n (\lambda_1, \ldots,
  \lambda_d) \in \bar{\Gamma}$, and where the approximation is up to
  $O(d \log(n) + \log(n/\eps))$.
\end{lemma}

The proof of Lemma~\ref{lem:Hqcalc} is based on the fact that there
exists a measurement on $\sigma_U[\lambda]^{\otimes n}$ such that the
resulting state, conditioned on a certain measurement outcome, is
equal to the state $\rho_{\bU}^n$. The assertion then follows from the
observation that this measurement does only change the entropies by a
small constant.

We start with the proof of a restricted version of the statement,
formulated as Lemma~\ref{lem:asym} below, which holds for states of
the form~\eqref{eq:rhoU} where only one of the weights $\mu_{\bn}$ is
nonzero. Let $\ket{\varphi_1}, \ldots, \ket{\varphi_d} \in \cH_U
\otimes \cH_W$ be purifications of the states $\sigma_1, \ldots,
\sigma_d$, respectively, such that the partial traces
$\tr_{U}(P_{\ket{\varphi_i}})$ are mutually orthogonal.  For $\bn =
(n_1, \ldots, n_d) \in \Gamma^n_d$, let
\begin{equation} \label{eq:psiUW}
    \ket{\psi}_{\bU \bW}^\bn
  :=
    \frac{1}{\sqrt{|S_n|}} \sum_{\pi \in S_n} \pi\bigl(
      \ket{\varphi_1}^{\otimes n_1} \otimes \cdots
        \otimes \ket{\varphi_d}^{\otimes n_d}
     \bigr) \ ,
\end{equation}
where $S_n$ denotes the set of all permutations $\pi$ on $n$-tuples.
Similarly, for $\lambda = (\lambda_1, \ldots, \lambda_d)$, let
\begin{equation} \label{eq:phiUW}
  \ket{\varphi}_{U W}^\lambda
:=
  \sum_{i=1}^d \sqrt{\lambda_i} \, \ket{\varphi_i} \ .
\end{equation}

\begin{lemma} \label{lem:asym}
  Let $\rho_{\bU \bW}^n[\bn] := P_{\ket{\psi}_{\bU \bW}^\bn}$ be the
  pure state defined by~\eqref{eq:psiUW}, for some fixed $\bn = (n_1,
  \ldots, n_d) \in \Gamma^n_d$, and let $\rho_{\bV \bW}^n[\bn] := (\cE
  \otimes \one_W)^{\otimes n}(\rho_{\bU \bW}^n[\bn])$. Moreover, for
  $\lambda := (\frac{n_1}{n}, \ldots, \frac{n_d}{n})$, let $\sigma_{U
    W}[\lambda] := P_{\ket{\varphi}_{U W}^\lambda}$ be the pure state
  defined by~\eqref{eq:phiUW} and let $\sigma_{V W}[\lambda] := (\cE
  \otimes \one)(\sigma_{U W}[\lambda])$.  Then, for $\alpha \in
  \{0,2\}$,
  \[
    \bigl| \Hq_\alpha^\eps(\rho_{\bV \bW}^n[\bn])
    - n \Hq(\sigma_{V W}[\lambda]) \bigr|
  \leq
    O(\log(n/\eps)) \ .
  \]
\end{lemma}

\begin{proof}
  For any $i \in \{1, \ldots, d\}$, let $P_i$ be the projector onto
  the support of $(\cE \otimes \one_W)(P_{\ket{\varphi_i}})$, which,
  by the definition of the vectors $\ket{\varphi_i}$, are orthogonal
  for distinct $i$.  Additionally, let $\cF: \rho \mapsto F_0 \rho
  F_0^\dagger + F_1 \rho F_1^\dagger$ be the measurement on
  $\cH_V^{\otimes n}$ defined by
  \[
    F_0
  :=
    \sum_{\pi \in S_n}
      \pi(P_1^{\otimes n_1} \otimes \cdots \otimes P_d^{\otimes n_d})
  \]
  and $F_1 := \one - F_0$. We first show that
  \begin{equation} \label{eq:projeq}
    \rho_{\bV \bW}^n[\bn]
  =
    \frac{1}{N} F_0
      \bigl(\sigma_{V W}[\lambda]^{\otimes n}\bigr) F_0^{\dagger} \ ,
  \end{equation}
  where $N := |S_n| \prod_{i=1}^d \lambda_i^{n_i}$.
  
  Let $(\cE\otimes \one_W)(\rho) = \sum_{\alpha=1}^m \bar{E}_\alpha
  \rho \bar{E}_\alpha^\dagger$ be the operator-sum representation of
  $\cE \otimes \one_W$. Moreover, for any $\bar{\alpha}:=(\alpha_1,
  \ldots, \alpha_n)$, let $\bar{E}_{\bar{\alpha}} :=
  \bar{E}_{\alpha_1} \otimes \cdots \otimes \bar{E}_{\alpha_n}$.  The
  above equality can then be rewritten as
  \[
    \sum_{\bar{\alpha}}
      \bar{E}_{\bar{\alpha}}
      (\rho_{\bU \bW}^n[\bn])
      \bar{E}_{\bar{\alpha}}^\dagger \\
  =
    \frac{1}{N}
    \sum_{\bar{\alpha}}
      F_0 \bar{E}_{\bar{\alpha}}
        (\sigma_{U W}[\lambda]^{\otimes n})
          \bar{E}_{\bar{\alpha}}^\dagger
        F_0^\dagger \ .
  \]
  It suffices to verify that equality holds for any term in the sum,
  i.e.,
  \begin{equation} \label{eq:projtest}
      \bar{E}_{\bar{\alpha}} \ket{\psi}_{\bU \bW}^n
  =
    \frac{1}{\sqrt{N}}
      F_0 \bar{E}_{\bar{\alpha}}
        \ket{\varphi}_{\bU \bW}^{\otimes n} \ ,
  \end{equation}
  for any $n$-tuple $\bar{\alpha} = (\alpha_1, \ldots, \alpha_n)$ on
  $\{1, \ldots, m\}$. Because of the definition of the projectors
  $P_i$, we have $P_{i} \bar{E}_{\alpha} \ket{\varphi_j} =
  \bar{E}_{\alpha} \ket{\varphi_j}$, if $i = j$, and $P_{i}
  \bar{E}_{\alpha} \ket{\varphi_j} = 0$ otherwise.  Hence, for any
  $\ket{\varphi_{i_1, \ldots, i_n}} := \ket{\varphi_{i_1}} \otimes
  \cdots \otimes \ket{\varphi_{i_n}}$,
  \[
    F_0 \bar{E}_{\bar{\alpha}}
      \ket{\varphi_{i_1, \ldots, i_n}}
  =
    \begin{cases}
      \bar{E}_{\bar{\alpha}} \ket{\varphi_{i_1, \ldots, i_n}}
      & \text{if $\ket{\varphi_{i_1, \ldots, i_n}} \in \Theta_\bn$} \\
      0 & \text{otherwise,}
    \end{cases}
  \]
  where $\Theta_\bn := \{\pi(\ket{\varphi_1}^{n_1} \otimes \cdots
  \otimes \ket{\varphi_d}^{n_d}) : \pi \in S_n \}$. This
  implies~\eqref{eq:projtest} and thus~\eqref{eq:projeq}.

  Let $\rho_{\tilde{\bV} \tilde{\bW}}^n$ be the state of the system
  after applying the measurement $\cF$ to $\sigma_{V
    W}[\lambda]^{\otimes n}$, and let $Z$ be the classical measurement
  outcome.  In the following, we write $S_\alpha^\eps(\tilde{\bV}
  \tilde{\bW}|Z=0)$ to denote the entropy of $\rho_{\tilde{\bV}
    \tilde{\bW}}^n$ conditioned on $Z=0$.  Then, according
  to~\eqref{eq:projeq},
  \begin{equation} \label{eq:Scondeq}
    S_\alpha^\eps(\rho_{\bV \bW}^n[\bn])
  =
    S_\alpha^\eps(\tilde{\bV} \tilde{\bW}|Z=0) \ .
  \end{equation}
  Let $\eps':= \frac{1}{2}P_Z(0) \eps$ where $P_Z(0) = \tr(F_0
  (\sigma_{V W}^{\otimes n}) F_0^\dagger)$.
  Using~\eqref{eq:Hinfcondvalue} and~\eqref{eq:Hinfmeasupbound}, we
  find
  \[
  \begin{split}
    S_2^\eps(\tilde{\bV} \tilde{\bW}|Z=0)
  & \geq
    S_2^{2\eps'}(\tilde{\bV} \tilde{\bW} | Z) \\
  & >
    S_2^{\eps'}(\sigma_{V W}[\lambda]^{\otimes n}) - 1 - 2 \log(1/\eps') \ .
  \end{split}
  \]
  Similarly, using~\eqref{eq:Hzerocondvalue}
  and~\eqref{eq:Hzeromeaslowbound},
  \begin{multline*}
    S_0^\eps(\tilde{\bV} \tilde{\bW}|Z=0)
  \leq
    S_0^{2 \eps'}(\tilde{\bV} \tilde{\bW}^n|Z)  \\
  \leq
    S_0^{2 \eps'}(\sigma_{V W}[\lambda]^{\otimes n}) + 1 \ .
  \end{multline*}
  Hence, because the smooth R\'enyi entropy of order $0$ is larger
  than the smooth R\'enyi entropy of order $2$
  (cf.~\eqref{eq:Salphabeta}), we have
  \begin{multline*}
    S_2^{\eps'}(\sigma_{V W}^{\otimes n})
  \lessapprox
    S_2^\eps(\tilde{\bV} \tilde{\bW}|Z=0) \\
  \lessapprox
    S_0^\eps(\tilde{\bV} \tilde{\bW}|Z=0)
  \lessapprox
    S_0^{2 \eps'}(\sigma_{V W}^{\otimes n})
  \end{multline*}
  where the approximation holds up to $O(\log(1/\eps'))$.  Combining
  this with~\eqref{eq:Scondeq}, we conclude
  \[
    S_2^{\eps'}(\sigma_{V W}^{\otimes n})
  \lessapprox
    S^\eps_{\alpha}(\rho_{\bV \bW}^n[\bn])
  \lessapprox
    S_0^{2 \eps'}(\sigma_{V W}^{\otimes n}) \ .
  \]

  The assertion then follows from the observation that $P_Z(0) \geq
  \frac{1}{n}$, which implies $\eps' \geq \frac{\eps}{2n}$, and the
  fact that the smooth R\'enyi entropy of product states approaches
  the von Neumann entropy (see~\eqref{eq:typsec}).
\end{proof}

\begin{proof}[Proof of Lemma~\ref{lem:Hqcalc}]
  It is easy to see that it suffices to prove the assertion for one
  specific purification of the states $\rho_U^n$ and $\sigma_U$.  Let
  thus $\ket{\varphi_1}, \ldots, \ket{\varphi_d} \in \cH_U \otimes
  \cH_W$ be the purifications of $\sigma_1, \ldots, \sigma_d$ defined
  above.  Moreover, for any $\bn \in \Gamma^n_d$, let $\rho_{U
    W}^n[\bn] := P_{\ket{\psi}_{\bU \bW}^{\bn}}$ be the state defined
  by~\eqref{eq:psiUW} and let $\rho_{\bU \bW}^n := P_{\ket{\psi}}$ where
  \[
    \ket{\psi}
  :=
    \sum_{\bn \in \Gamma}
      \sqrt{\mu_{\bn}} \ket{\psi}_{\bU \bV}^{\bn} \ .
  \]
  Similarly, for any $\lambda = (\lambda_1, \ldots, \lambda_d)$, let
  $\sigma_{U W}[\lambda] := P_{\ket{\varphi}_{U W}^\lambda}$ be the
  state defined by~\eqref{eq:phiUW}.  It follows from these
  definitions that $\rho_{\bU \bW}^n$ is a purification of
  $\rho_{\bU}^n$ and, similarly, $\sigma_{U W}[\lambda]$ is a
  purification of $\rho_U[\lambda]$.

  For any $\bn \in \Gamma^n_d$, let $\cH_W^{\bn}$ be the smallest
  subspace of $\cH_W^{\otimes n}$ containing the support of the traces
  $\rho_{\bW}^n[\bn] = \tr_{\cH_U^{\otimes n}}(\rho_{\bU
    \bW}^n[\bn])$.  By the definition of the vectors
  $\ket{\varphi_i}$, the subspaces $\cH_W^{\bn}$ are orthogonal for
  distinct $\bn \in \Gamma^n_d$.  Hence, there exists a projective
  measurement $\cF$ onto the subspaces $\cH_U \otimes \cH^{\bn}_W$.
  Consider the state $\rho_{\tilde{\bV} \tilde{\bW}}$ obtained when
  applying $\cF$ to $\rho_{\bV \bW}^n$, and let $Z$ be the classical
  outcome, i.e., $Z$ takes values from the set $\Gamma^n_d$.  The
  entropy $S_\alpha^\eps(\tilde{V}^n \tilde{W}^n|Z=\bn)$ of the state
  $\rho_{\tilde{\bV} \tilde{\bW}}^n$ after the measurement,
  conditioned on $Z=\bn$, is equal to the entropy of $\rho_{\bV
    \bW}^n[\bn]$ as defined by Lemma~\ref{lem:asym}, i.e.,
  \[
    S_\alpha^\eps(\tilde{\bV} \tilde{\bW}|Z=\bn)
  =
    S_\alpha^\eps(\rho_{\bV \bW}^n[\bn]) \ .
  \]
  Hence, from~\eqref{eq:Hinfmeaslowbound}
  and~\eqref{eq:Hinfeventbound},
  \[
  \begin{split}
    S_2^\eps(\rho_{\bV \bW}^n)
  & \geq
    S_2^{\eps}(\tilde{\bV} \tilde{\bW}|Z) - \Hc_0(Z) \\
  & \geq
    \min_{\bn \in \bar{\Gamma}} S_2^{\eps/2}(\tilde{\bV} \tilde{\bW}|Z=\bn) -
  \Hc_0(Z) \\
  & =
    \min_{\bn \in \bar{\Gamma}} S_2^{\eps/2}(\rho_{\bV \bW}^n[\bn]) - \Hc_0(Z) \ .
  \end{split}
  \]
  and, similarly, from \eqref{eq:Hzeromeasupbound}
  and~\eqref{eq:Hzeroeventbound},
  \[
  \begin{split}
    S_0^\eps(\rho_{\bV \bW}^n)
  & \leq
    S_0^\eps(\tilde{\bV} \tilde{\bW}|Z) + \Hc_0(Z) \\
  & \leq
    \max_{\bn \in \bar{\Gamma}} S_0^{\eps/2}(\rho_{\bV \bW}^n[\bn])
    + \Hc_0(Z) \ .
  \end{split}
  \]

  Finally, from Lemma~\ref{lem:asym},
  \[
    \bigl| S_\alpha^{\eps/2}(\rho_{\bV \bW}^n[\bn])
           - n S_\alpha(\sigma_{V W}[\lambda]) \bigr|
  \leq
    O(\log(2 n/\eps)) \
  \]
  where $\lambda = (\frac{n_1}{n}, \ldots, \frac{n_d}{n})$.  The
  assertion then follows from the observation that $\Hc_0(Z) \leq
  \log_2(|\Gamma^n_d|) \leq d \log_2(n)$.
\end{proof}

\section{Entropy of almost product states} 

Let $X$ be a classical random variable and let $\rho_B^x$ be a quantum
state depending on $X$. Clearly, if the states $\rho_B^x$ are equal
for all $x$, then the entropy of $X$ does not change when conditioning
on the quantum system, i.e., $S(X) = S(X|B)$. In this section, we show
that, if the joint state describing $X$ and $\rho_B^x$ is close to a
product state, then the entropy change of $X$ when conditioning on the
quantum system is still small (cf.\ Lemma~\ref{lem:SABbound}).

We first need a lemma relating the trace distance of two density
operators to the trace distance of purifications of them.

\begin{lemma} \label{lem:puredist}
  Let $\rho$ and $\rho'$ be density operators and let $\ket{\psi}$ be
  a purification of $\rho$. Then there exists a purification
  $\ket{\psi'}$ of $\rho'$ such that
  \[
    \dist(P_{\ket{\psi}}, P_{\ket{\psi'}})
  \leq
    \sqrt{2 \dist(\rho, \rho')} \ .
  \]
\end{lemma}

\begin{proof}
  Note that the fidelity $F$ is related to the trace distance $\dist$
  according to
  \[
    1-F(\rho,\sigma) \leq \dist(\rho,\sigma) \leq \sqrt{1-F(\rho,\sigma)^2} \ .
  \]
  Moreover, Uhlmann's theorem states that there exists a purification
  $\ket{\psi'}$ of $\rho'$ such that
  \[
  F(\rho, \rho') = F(P_{\ket{\psi}}, P_{\ket{\psi'}}) \ .
  \]
  Hence,
  \[
  \begin{split}
    \dist(P_{\ket{\psi}}, P_{\ket{\psi'}})
  & \leq
    \sqrt{1-F(P_{\ket{\psi}}, P_{\ket{\psi'}})^2} \\
  & =
    \sqrt{1-F(\rho, \rho')^2} \\
  & \leq
    \sqrt{2 (1-F(\rho, \rho'))} \\
  & \leq
    \sqrt{2 \dist(\rho, \rho')} \ .
  \end{split}
  \]
\end{proof}

\begin{lemma} \label{lem:SABbound}
  Let $\rho_{X B}$ be a bipartite density operator of the form
  \[
    \rho_{X B}
  =
    \sum_{x=1}^d \mu_x P_{\ket{x}} \otimes \rho^x_B \ ,
  \]
  where $\{\ket{x}\}_{x \in \{1, \ldots, d\}}$ is an orthonormal
  basis of the first subsystem. If
  \[
    \dist(\rho_{X B}, \rho_{X} \otimes \rho_{B})
  \leq
    \eps \ ,
  \]
  then
  \[
    S(X|B) \geq S(X) - \sqrt{2 \eps} \log(d) - 1/e \ .
  \]
\end{lemma}

\newcommand*{\Exp}{E}

\begin{proof}
  It is easy to see that the trace distance between $\rho_{X B}$ and
  $\rho_X \otimes \rho_B$ can be written as
  \[
    \dist(\rho_{X B}, \rho_{X} \otimes \rho_{B})
  =
    \sum_x \mu_x \bigl(\dist(\rho_B^x, \rho_B)\bigr) \ .
  \]
  Let $\psi$ be a purification of $\rho_B$. According to
  Lemma~\ref{lem:puredist}, for all $x \in \{1, \ldots, d\}$, there
  exists a purification $\ket{\psi_x}$ of $\rho_B^x$ such that
  \[
    \dist(P_{\ket{\psi_x}}, P_{\ket{\psi}})
  \leq
    \sqrt{2 \dist(\rho^x_B, \rho_B)} \ .
  \]
  Hence, using Jensen's inequality,
  \begin{multline*}
    \sum_{x} \mu_x \bigl(\dist(P_{\ket{\psi_x}},
  P_{\ket{\psi}})\bigr) 
  \leq
    \sqrt{2 \sum_{x} \mu_x \bigl(\dist(\rho_B^x, \rho_B)\bigr)}
  \leq
    \sqrt{2 \eps} \ .
  \end{multline*}
  Let now $\rho_{X B B'}$ be the state defined by
  \[
    \rho_{X B B'}
  :=
    \sum_{x} \mu_x \bigl(P_{\ket{x}} \otimes P_{\ket{\psi_x}}\bigr) \ .
  \]
  Note that, by this definition, $\rho_{X B} = \tr_{B'}(\rho_{X B
    B'})$.

  From the strong subadditivity, we have
  \begin{multline*}
    S(X|B)
  \geq
    S(X|B B') \\
  =
    S(X B B') - S(B B')
  \geq
    S(X) - S(B B')
  \end{multline*}
  where the last inequality holds since
  \[
    S(B B'|X) = \sum_{x} \mu_x S(\rho_x^{B B'}) \geq 0 \ .
  \]
  Because the rank of $\rho_{B B'}$ is not larger than $d$, $S(B B')$
  can be bounded using Fannes' inequality, i.e.,
  \begin{equation} \label{eq:SBBbound}
    S(\rho_{B B'})
  \leq
    S(P_{\ket{\psi}}) + \dist(\rho_{B B'}, P_{\ket{\psi}}) \log(d) + 1/e \ .
  \end{equation}
  Since $\rho_{B B'} = \sum_{x} \mu_x (P_{\ket{\psi_x}})$, it follows from
  the convexity of the trace distance that
  \[
    \dist(\rho_{B B'}, P_{\ket{\psi}})
  \leq
    \sum_{x} \mu_x \bigl(\dist(P_{\ket{\psi_x}}, P_{\ket{\psi}}) \bigr)
  \leq
    \sqrt{2 \eps} \ .
  \]
  Inserting this into~\eqref{eq:SBBbound} and observing that
  $S(P_{\ket{\psi}}) = 0$ concludes the proof.
\end{proof}

\section{Known results} \label{sec:known}

Consider two different measurement operations $\cE$ and $\cF$ applied
to the individual parts of a symmetric state $\rho^n$.
Lemma~\ref{lem:Ekert} gives a relation between the measurement
statistics of $\cE$ and $\cF$ (see~\cite{ChReEk04} for a proof).

\begin{lemma} \label{lem:Ekert}
  Let $\rho^n$ be a symmetric quantum state on $\cH^{\otimes n}$, and
  let $\cE$ and $\cF$ be POVMs on $\cH$ with $|\cE|$ and $|\cF|$ POVM
  elements, respectively. Let $Q_{\bX}$ and $Q_{\bY}$ be the frequency
  distribution of the outcomes when applying the measurements
  $\cE^{\otimes k}$ and $\cF^{\otimes n-k}$, respectively, to
  different subsystems of $\rho^n$.  Finally, let $\cB$ be any convex
  set of density operators such that, for any operator $A$ on $n-1$
  subsystems, the normalization of $\tr_{n-1}(\one \otimes A \rho^n
  \one \otimes A^\dagger)$ is contained in $\cB$.  Then, for any $\eps
  > 0$, with probability at least $1-2^{|\cE| + |\cF|} e^{-\frac{n
      \eps^2}{8}}$, there exists a state $\sigma \in \cB$ such that
  \[
    \frac{k}{n} \dist\bigl(Q_{\bX}, P_\cE[\sigma]\bigr)
    + \frac{n-k}{n} \dist\bigl(Q_{\bY}, P_\cF[\sigma]\bigr) \leq \eps \ ,
  \]
  where $P_\cE[\sigma]$ and $P_\cF[\sigma]$ denote the probability
  distributions of the outcomes when measuring $\sigma$ with respect
  to $\cE$ and $\cF$, respectively.
\end{lemma}

Lemma~\ref{lem:pa} below provides an expression for the maximum length
of a key $S$ that can be generated from a string $Z$ such that $S$ is
secure against an adversary holding a quantum state $\rho_E^z$
depending on $Z$.  The proof can be found in~\cite{RenKoe05} (see
also~\cite{KoMaRe03}). Note that Lemma~\ref{lem:pa} holds with respect
to a so-called \emph{universally composable} security definition.
This implies that the final key $S$ can be used in \emph{any} context
where a perfect key (i.e., a uniformly distributed key which is
completely independent of the adversary's knowledge) is secure.

\begin{lemma} \label{lem:pa}
  Let $\rho_{Z E}$ be a density operator such that $\rho_Z$ is
  classical, i.e., $\rho_{Z E} = \sum_{z} P_Z(z) P_{\ket{z}} \otimes
  \rho_E^z$, where $\{\ket{z}\}_z$ is a family of orthonormal vectors,
  and let $\eps \in \bbR^+$. Let $S$ be the key computed by applying a
  two-universal hash function $F$ mapping the value of $Z$ to a value
  in $\{0,1\}^\ell$. Then $S$ is $\eps$-secure with respect to
  $\rho_{E F}$ if
  \[
    \ell
  \leq
    S_{2}^{\eps'}(Z E)
    - S_0^{\eps'}(E)
    - 2 \log(1/\eps) \ ,
  \]
  where $\eps' = (\eps/8)^2$.
\end{lemma}

The following lemma on error correction is a direct consequence of
Lemma~4 from~\cite{RenWol04b} (see also~\cite{RenWol04}). Roughly
speaking, it states that a message of length $\Hc_0^{\eps}(X|Y)$ is
sufficient to guess the value of $X$ when only $Y$ is known.

\newcommand*{\cR}{\mathcal{R}}

\begin{lemma} \label{lem:ir}
  Let $\cX$ and $\cY$ be sets, let $\eps \in \bbR^+$, and let $m \in
  \bbN$. Then there exists a probabilistic encoding function $e: \cX
  \times \cR \rightarrow \cC$, taking randomness with some
  distribution $P_R$ such that the following holds: For all
  probability distributions $P_{X Y}$ on $\cX \times \cY$ satisfying
  $\Hc_0^{\eps'}(X|Y) + \log(1/\eps') \leq m$, for $\eps'=\eps/2$,
  there exists a decoding function $d: \cC \times \cY \rightarrow \cX$
  such that the probability of a decoding error is smaller than
  $\eps$, i.e.,
  \[
    \Pr_{(x,y,r) \leftarrow P_{X Y} \times P_R}
      \bigl[d(e(x,r), y) = x\bigr]
  \geq
    1-\eps
  \]
  and the encoding $C:=e(X,R)$ gives no more than $m$ bits of
  information on $X$, i.e.,
  \[
  \Hc_0(C) - \Hc_\infty(C|X) \leq m \ .
  \]
\end{lemma}

\end{document}